\documentclass[12pt]{article}
\usepackage{epsfig,latexsym}

\input epsf

\def \N {$\mathcal{N}$}
\def \nc {$N_c$}
\def \k {\kappa} 
\def \a {C_a}
\def \b {C_{\tilde{a}}}
\def \g {C_l}
\def \P {\Phi}
\def \ep{\epsilon}
\def \t {\tau}
\def\p{\partial}

\def\Tr{{\rm Tr}}

\def\nn{\nonumber}

\def \C {{\cal C}}

\begin{document}
\thispagestyle{empty}
\font\cmss=cmss10 \font\cmsss=cmss10 at 7pt
\par\hfill Bicocca-FT-03-02
\par\hfill IFUP-TH 8/2003\\
\vskip .1in \hfill hep-th/0301118

\hfill

\vspace{20pt}

\begin{center}
{\Large \textbf{Non Supersymmetric Regular Solutions from Wrapped and Fractional Branes}}
\end{center}

\vspace{6pt}

\begin{center}
\textbf{Riccardo Apreda 
}\\ \vspace{20pt}
\bigskip
\textit{
Dipartimento di Fisica, Universit\`{a} di Pisa,}\\ 
{\small via Buonarroti, 2; I-56127 Pisa, Italy.}\\
\texttt{apreda@df.unipi.it}\\[6mm]

\end{center}

\vspace{12pt}

\begin{center}
\textbf{Abstract}\\
We present two classes of regular supergravity backgrounds dual to supersymmetric and non-supersymmetric gauge theories living on the world-volume of wrapped branes.\\
In particular we consider the Maldacena Nu\~nez and the Klebanov Strassler models, 
describing \N=1 and \N =2 theories, 
and find their non supersymmetric generalization by explicitly solving the second order equation of motion.
We also study various aspects of these solutions, including the supersymmetry breaking issue and the vacuum energy.
\end{center} 
\vfill

\newpage
\section{Introduction}
%
The main goal of the gravity/field theory duality is to extend the original AdS/CFT correspondence to more realistic gauge theories like QCD, non conformal and with broken or absent supersymmetry.
Several approaches have been used. For example one can deform the CFT by adding  finite temperature corrections \cite{finitetemp}. Another possibility is to use directly non supersymmetric string theories \cite{type0}.

Moreover one could start from a supersymmetric background and then deform it in order to break SUSY on the field theory side of the correspondence.
An advantage of this approach is that it allows to study also \N =2 and \N =1 theories, which of course are interesting by themselves.

Deformations that leave the spacetime asymptotically AdS are obtained adding relevant operators which induce a RG flow in the field theory living on the brane \cite{LPPZ,GF,PS}.

As concern geometries which are not even asymptotically AdS,
one possible deformation consists of putting branes on singular backgrounds, so that the singularity alters the inner-space symmetry group. Another way to break some symmetry is to wrap  branes around compact surfaces. Various dispositions of intersecting branes are finally useful for the description of \N =2 theories.
However all these methods usually leads to some kind of singularity of the supergravity background.
In general stringy corrections are assumed to resolve singularities of the supergravity approximation (for the \N=2 model of \cite{enhancon} this seems indeed to be the case, via the so called enhan\c con mechanism).
Nevertheless two cases admitting globally regular solutions already at supergravity level are known.

The first was introduced by Maldacena and Nu\~nez (MN) in \cite{MaNu},
and consists of $N$\ \ NS5-branes wrapped on an $S^2$. 
The other case was introduced by Klebanov and Strassler (KS) in \cite{KS}, and consists of a set of N regular D3 branes and M fractional D3 branes placed on a conifold singularity. 
These systems seem quite different at first glance, but actually the local geometry of the two non singular solutions is very similar.
These models were studied in several papers, mainly in the context of preserved SUSY, so that first order differential equations could be used (see \cite{papersonMN} for related papers not cited elsewhere). 

In \cite{gubser1} the authors solve directly, in a numerical way, the
second order equation of motion of the MN set-up, and find a class of non extremal solutions, with only one scalar deformation added to the original case.
In the first part of this paper we generalize the results of \cite{gubser1} and find a whole class of solutions with all the four interesting scalars switched on.
We also analyze various aspects of the relationships between the solutions we have found out and the corresponding gauge field theories.
We mention here that a linearized version of the equations of motion was used in \cite{ahason} to find approximated solutions with three scalars turned on. 

In the second part of the paper we solve directly the second order equation of motion for the other relevant  set-up, the KS one, again finding all possible interesting solutions (even if numerically). 
While this work was in preparation another work appeared \cite{gubser2} which also deals with non-supersymmetric extensions of the KS model. However the authors of that paper propose a different method of solution, based on a modification of the first order supersymmetric equations, and the class of solutions they find is quite different from our.
Even for the KS case we discuss various aspects of the correspondence between the field theories and our solutions.
\section{Gravity duals of gauge theories from wrapped five-
branes}
\subsection{Branes setup}
We consider the general setting of wrapped five-branes with world-volume $R^4 \times S^2$, which can be easily adapted to describe both \N=2, \N=1 and \N=0 theories.

In the case of \nc \ \ NS5 branes the world-volume theory is little string theory, which reduces in the IR to \N=1 six-dimensional SYM theory.
Wrapping two dimensions the resulting low-energy theory becomes effectively four dimensional.\footnote{Actually we do not get a ``pure'' - i.e. without matter - four dimensional (super) Yang Mills theory, since Kaluza Klein modes are not completely decoupled} Its gauge coupling is fixed by the volume of the wrapped dimensions, and a dependence of this volume on the transverse radius should be read as a running of the coupling.
The $SO(4) \sim SU(2)_+ \times SU(2)_-$ isometry group of the $R^4$ transverse space acts on the gauge theory as an $SO(4)$ R-symmetry;
on the other side of the correspondence, the relevant fields of the supergravity approximation dual to the field theory
are described by an $SO(4)_R$ gauged SUGRA.

The number of SUSYs inherited by the four dimensional theory depends on the wrapping surface. As well known however there are no covariantly constant spinor on $S^2$. To preserve some supersymmetry it is necessary to ``twist'' the wrapping of the branes on the surface along with the transverse radius, that is to twist the normal bundle 
. This operation corresponds in supergravity
to turning on abelian background fields in the $SO(4)$ group that cancel the spin connection of $S^2$.\\
From the schematic formula for the variation of a fermion:
\begin{equation}
\delta \Psi\sim D_{\mu}\epsilon=(\partial_{\mu}+\omega_{\mu}^{\nu\rho}\gamma^{\nu\rho}
-A_{\mu}^{ij}\Gamma^{ij})\epsilon,
\label{twist0}
\end{equation}
where Latin letters $i=1,2,3,4$ label internal $SO(4)$ indices,
we see that the surviving spinors are those satisfying the twist condition:
\begin{equation}
(\omega_{\mu}^{\nu\rho}\gamma^{\nu\rho}
-A_{\mu}^{ij}\Gamma^{ij})\epsilon=0.
\label{twist}
\end{equation}
The fields needed for the twist
are those of the decomposition
$SO(4)_R\rightarrow U(1)_{(1)}\times U(1)_{(2)}$, $A_{(1)}$
and $A_{(2)}$.
To get $\mathcal{N}=2$ solution
only $A_{(1)}$ is turned on,
while in the $\mathcal{N}=1$ solution both $A_{(1)}$ and $A_{(2)}$ are turned on.

However in the presence only of these abelian $A$ fields, the background is still singular. Finally in the MN model the switching on of the full non abelian field resolves the singularity.\\[3mm]
\subsection{The supergravity action}
Five-branes are soliton of type IIB Supergravity in ten dimensions.
In MN setup however besides the world volume coordinates only the radius of the transverse space is relevant, while the other three coordinates do not play any role. It is thus convenient to use the consistent reduction  down to seven dimensional, $SO(4)$ gauged supergravity by the Kaluza-Klein mechanism as in \cite{cvetic1}.
Actually the same set up can by derived from M-theory, where the wrapping on an $S_2$ of \nc \ \ M5 branes is well described by the approximation of seven dimensional $SO(5)$ gauged SUGRA; type IIB theory and $SO(4)$ gauged SUGRA are obtained with the singular limit described in \cite{cvetic1}.
The seven-dimensional solutions can be then lifted up to ten dimension again using the method of \cite{cvetic1,cham1}.\\
The relevant terms of the bosonic lagrangian of $SO(4)$ gauged SUGRA are \cite{sesa}:
\begin{equation}
2\kappa^2 {\cal L} = \sqrt{-g} \left( R + {1\over 2} m^2 (T^2 - 2T_{ij}T^{ij}) -
Tr(P_{\mu} P^{\mu}) - {1\over 2} (V_I{}^i V_J{}^j F_{\mu \nu}^{IJ})^2 \right) . \label{lagmanu}
\end{equation}
As we have seen in the previous section, the gauge fields in $F_{\mu\nu}^{IJ}$ parameterize the twist of the normal bundle.
We take the $SO(4)=SU(2)^+ \times SU(2)^-$ gauge fields of the form:
\begin{eqnarray}
A & = & \alpha \,[\cos{\theta} \, d\phi \, \eta _{1}^{+} + a(\rho) \,d\theta \, \eta
_{2}^{+}+ b(\rho) \, \sin{\theta} \, d\phi \, \eta _{3}^{+}]+ \nonumber \\
&& \beta \, [\cos{\theta}
\, d\phi \, \eta _{1}^{-} + \tilde{a}(\rho) \, d\theta \, \eta _{2}^{-} +
\tilde{b}(\rho) \, \sin{\theta} \, d\phi \, \eta _{3}^{-}] \label{AA}.
\end{eqnarray}
The field strength is normalized as $F=dA+2m[A,A]\,$;
the $\eta$ matrices are the generators of the $SU(2)^{\pm}$ in
the $SO(4)$ notation and take the form:
\begin{eqnarray}
\eta _{1}^{\pm}={1\over 2} \texttt{\footnotesize $\left( \begin{array}{cccc} 0 & 1 & 0 & 0
\\ -1 & 0 & 0 & 0 \\ 0 & 0 & 0 & \pm 1 \\ 0 & 0 & \mp 1 & 0 \end{array}
\right)$}\,\, \eta_{2}^{\pm}={1\over 2}  \texttt{\footnotesize $
\left( \begin{array}{cccc} 0 & 0 & \mp 1 & 0 \\ 0 & 0 & 0 & 1 \\
\pm 1 & 0 & 0 & 0 \\ 0 & -1 & 0 & 0
\end{array} \right)$}\,\,
\eta _{3}^{\pm}={1\over 2}  \texttt{\footnotesize $\left(\begin{array}{cccc} 0 & 0 & 0 & 1 \\
0 & 0 & \pm 1 & 0 \\ 0 & \mp 1 & 0 & 0 \\ -1 & 0 & 0 & 0 \end{array}
\right)$}
\label{eta}
\end{eqnarray}
We remember that the fields $A_{(i)}$ in the end of the previous section are defined as:
\[
A_{(1)}=\frac{1}{2} \cos \theta \; d\phi\; (\eta_+ + \eta_-); \qquad \qquad
A_{(2)}=\frac{1}{2} \cos \theta\; d\phi \; (\eta_+ - \eta_-).
\]
$V_i^I$ is the symmetric matrix for the 10 scalar degrees of freedom
parameterizing the $SL(4,R)/SO(4)$ coset space ($I$ and $i$ are respectively the gauge and composite $SO(4)$
indices). Their values give the position of the branes in the transverse space.
The $T$ matrix is
defined as $T_{ij} = V_i^{-1\,I}
V_j^{-1\,J} \delta_{IJ}$,  $T=T_{ij}\delta_{ij}$.
 The kinetic term for the scalars $P_{\mu}$ is the symmetric part of $V_i^{-1\,I} {\cal D}_{\mu} V_I{}^j = \left( Q_{\mu} \right)_{[ij]} + \left( P_{\mu}
\right)_{(ij)}$, where the
covariant derivatives are defined as ${\cal D}_\mu V_I{}^j=\partial_\mu
V_I{}^j+2m A_{\mu\, I}^J V_J{}^j$ on the scalars, ${\cal D}_\mu \psi=(\partial_{\mu}+{1 \over 4}Q_{\mu ij}\Gamma^{ij}+{1 \over 4}\omega_{\mu}^{\nu\lambda}\gamma^{\nu\lambda})\psi$ on the spinors.

Before twisting the ten scalar are dual to the bilinear operator $Tr X_i X_j$, where $X_i$ are the six dimensional scalar fields. Upon compactification on $S^2$ and related twist, $X_1$ and $X_2$ get masses. From a geometrical point of view it means that the branes are no more free to move in the two twisted directions.  In the \N=2 case $X_3$ and $X_4$ remain massless, and parameterize the moduli space of \N=2 field theories. 
Thus the relevant solutions are expected to involve only the three scalars left (one being the dilaton) and the matrix $V_{iJ}$ can be diagonalized to the form:
\begin{equation}
V_I{}^i = {\rm diag} e^{\frac{f}{2}}( e^{-\frac{l}{2}},  e^{-\frac{l}{2}},
e^{\frac{l}{2}-k},  e^{\frac{l}{2}+k}
).
\end{equation}
In the \N=1 case also  $X_3$ and $X_4$ get masses from the twist (there is no moduli space for \N=1), but it is still possible to use the above general form for the matrix  $V_{iJ}$, and constraint it later on.\\
The ansatz for the metric (in the Einstein frame) is:
\begin{equation}
ds_{7}^2 = e^{2f}(dx_{4}^2 + d\rho^2) + e^{2g}(d\theta^2 + \sin^{2}{\theta} d\phi^2).
\label{m2}
\end{equation}
Moreover the integration of the action in the variable $\theta$ selects $b(\rho) = a(\rho)$; for $\beta \neq 0$ this also leads to $\tilde{b}= \tilde{a}$ and $\beta = \alpha$, while for $\beta=0$ (\N=1) it is necessary to put $k(\rho)$=0.

Finally for the full $SO(4)$, choosing the normalization $\alpha=\beta=\frac{1}{2},$
$\; \; m=1$ and the notation $G(\rho)=exp[2 (g - f)]$, the resulting unidimensional effective lagrangian can be written:
\[
\mathcal{L} = e^{5 f} G^{-1}  \left(  \frac{\dot{G}^2}{2} + 25 \dot{f}^2 G^2+10 G \dot{G}\dot{f}- \dot{l}^2 G^2-2 \dot{k}^2 G^2-\frac{ G}{2} (\dot{a}^2+\dot{\tilde{a}}) \cosh2k  - G \dot{a} \dot{\tilde{a}} \sinh2k\right.
\]
\begin{equation}
+2 G+G^2 (4 \cosh2k-2 e^{-2 l} \sinh2k^2)- G (a^2+\tilde{a}^2) (\cosh2k \cosh2l-1) \label{lagmanu} \\[2mm]
\end{equation}
\[
\left.   -\frac{(a^2-1)^2+(\tilde{a}^2-1)^2}{4} \cosh2l-2 G a \tilde{a} \sinh2k \sinh2l+\frac{(a^2-1)(\tilde{a}^2-1)}{2} \sinh2l \right) 
\]
The $SU(2)_+$ lagrangian reduces to
\begin{eqnarray}
\mathcal{L} = e^{5 f} G^{-1}  \left(  \frac{\dot{G}^2}{2} + 25 \dot{f}^2 G^2+10 G \dot{G}\dot{f}- \dot{l}^2 G^2-\frac{ G}{2} \dot{a}^2 +2 G+4 G^2 \right.  \\ \left. - G a^2 (\cosh2l-1) -\frac{(a^2-1)^2}{4} \cosh2l \right) \nn \label{lagmanu2}
\end{eqnarray}
Starting from a different ansatz for the fields a particular case of these lagrangians was found in \cite{patsy,gubser1}.

The Eulero-Lagrange equations for the effective actions  coincide with the independent relationships in the full ten dimensional equations of motion.
We note however that the dimensional reduction is consistent only if all massive modes are effectively decoupled. Sometimes this is guaranteed by the symmetries of the problem. Otherwise the issue should be analyzed according to circumstances, and eventually checked by hand.

\subsection{Gauge theory: identification of the fields}
Every $SO(4)$ supergravity background is
dual to a deformation of the four dimensional $N=2$ SYM theory,
which has, besides the gauge field, two fermionic and two scalar
fields.

The two fermions, $\lambda$ and $\psi$, can be identified looking
at their charges under the various symmetries; in particular they
have opposite chirality and different behavior under the
R-symmetry generators.

Following \cite{abcpz},
we start from the original six dimensional
theory living on the five-brane, that is the (1,1) SYM. It has
fermions $\Psi=\Psi^+ + \Psi^-$, which transform in the
representation (4,2)+(4',2') of $SO(5,1)\times SO(4)_R$, and have
opposite (six-dimensional) chirality: $\Psi^{\pm}=\pm \gamma_7
\Psi^{\pm}$.

%
Since after wrapping the branes on the $S_2$ the Lorentz group is broken down to $SO(3,1)$ it is convenient to re-write
 these fermions as $ 2 \times 2 $ matrix of four dimensional spinors:
\begin{eqnarray}
\Psi^+=\left( \begin{array}{cc}
p & q \\ iq^c & -ip^c \end{array} \right), \qquad \Psi^-=\left(
\begin{array}{cc} \tilde{p} & \tilde{q} \\ i\tilde{q}^c &
-i\tilde{p}^c \end{array} \right).\label{psi}
\end{eqnarray}
%
Translating in this four dimensional notation the generators of the various global symmetries and imposing the consequent constraints on $\Psi$, we easily get 
the following charge assignments:
\vskip 0.5truecm
\begin{center}
{\small
\begin{tabular}{ccc|cccccccccccc}
 & & & & & & & & & &   \\[-1mm]
 & & & & $p=\lambda$ &  & $\tilde{p}=\bar{\psi}$ &  & $q$ &  & $\tilde{q}$  \\[-1mm]
 & & & & & & & & & &  \\[-1mm]
\hline
 & & & & & & & & & &   \\[-1mm]
 & $U(1)_R=U(1)_{(2)}$ & & & 1 &  & -1 &  & -1 &  & 1  \\[-1mm]
 & & & & & & & & & & & & & &  \\[-1mm]
 & $U(1)_J=U(1)_{S^2}$ & & & 1 &  & 1 &  & 1 &  & 1  \\[-1mm]
 & & & & & & & & & &  \\[-1mm]
 & $U(1)_{(1)}$ & & & 1 &  & 1 &  & -1 &  & -1  \\[-1mm]
 & & & & & & & & & &   \\
\end{tabular}}
\vskip 0.3truecm
Table 1: Charge assignment of the spinors.
\end{center}
\vskip 0.5truecm
Since $q$ and $\tilde{q}$ have opposite charges under the twisted $U(1)$ and under the $S_2$ generator,
they become massive after compactification.
Thus the two gauginos of four dimensional \N=2 SYM should be identified with the other two fermions remaining massless.
From the analysis of charges we conclude that $p$ has to be identified with $\lambda$ and $\tilde{p}$ with $\bar{\psi}$.

Now we can associate two particular bilinears of these fermions with the corresponding supergravity fields,
simply looking at the coupling:
\[
A^\mu_{ij}\bar{\Psi}\gamma^\mu\Gamma^{ij}\Psi.
\]
Substituting in this formula the expressions (\ref{AA}) for $A$
 the spinors $q$ and $\tilde{q}$ get opposite contributions in the sum over $\theta$ and $\phi$ indices and cancel
 out,while
 for the other fields we get that $a$ and $\tilde a$ couple
to the fermionic bilinears as:
\[
a\,\bar{\lambda^c}\lambda, \qquad \tilde{a}\bar{\psi^c}\psi.
\]
We remark here the fact that the same supergravity fields which
de-singularize the supergravity solution are on one hand
associated to the gaugino condensates, i.e. to the non trivial IR
dynamics of the gauge theory, on the other hand enter in the
process of soft supersymmetry breaking.

For the $N=1$ case one can repeat the same analysis and find again that the supergravity  field $a$ correspond
to the bilinear describing the gaugino condensate. Its behavior is in agreement  with field theory prediction, as already studied in \cite{abcpz,dvlm}\\
The identification for scalar fields is quicker.\\
The \N=2 SYM has two massless scalar degrees of freedom, encoded in the complex field $\phi$; its classical solutions give the two-dimensional moduli space of the theory. 

On the supergravity side, we have seen that the $T_{ij}$
matrix is related to the motion of the branes in the transverse space; every static disposition of the branes in the transverse space correspond to a vacua of the four dimensional theory. After the twist branes are no longer free to move in the twisted $R^2$, and every displacement in that directions correspond to an highly massive mode of field theory. Conversely displacements in the two untwisted dimensions
still have no energy cost and correspond to the massless scalars of the field theory, $i.e.$ to $\phi$; the freedom to choose the static position of branes finally gives rise to the moduli space of \N=2 SYM. 

The motion in the untwisted $R^2$ is parameterized by the sector
with $i,j=3,4$ only of  $T_{ij}$, and the corresponding scalars are dual to the
chiral bilinear operators of $N=2$ gauge theory. Thus the field
$l$, which parameterize the difference between the twisted and the
untwisted sector of $T_{ij}$, is associated to $Tr \phi
\bar{\phi}$, while field $k$ which explicitly breaks the remaining
$U(1)_R$ has to be associated with the operator $Tr \phi^2$.

In \N=1 SYM there are no massless scalars and no moduli space, and indeed with the different twist  all the modes describing branes displacements  get masses.
In the supergravity background the field $l$ can be still introduced, but there is no place for $k$, as we have seen in the previous section.\\
The dual of such backgrounds with $l$ turned on
are \N=1 field theories with a massive deformation, deformation that can be seen as the remainder of the breaking of an original \N=2 theory. \\
We stress the fact that while in the \N=2 case $l$ and $k$ correspond to operators which give vev or masses to massless fields, in the \N=1 case the leading perturbation associated to $l$ corresponds to adding mass to a field which is already very massive (as in \cite{ahason}).
\subsection{Supersymmetric solutions }
We review now the known analytic solutions, which are all supersymmetric.\\
The naive \N=1 solution has only one scalar in the coset space and only the abelian field turned on ($a(\rho)=0, \beta=0$ in (\ref{AA}) ).
This leads to the singular background:
\begin{equation}
G(\rho)= \rho, \qquad e^{-5 f(\rho)}=\frac{1}{\sqrt{\rho}}e^{\rho}, \qquad a(\rho)=0.
\end{equation}
The non-singular, \N=1 solution of \cite{MaNu}, corresponding to one scalar in the coset space but group $SU(2)^+$, can be obtained from the general equations setting $l=0, k=0$ and $\beta=0$.
The resulting equations are solved by the functions in \cite{MaNu}:
\begin{equation}
G(\rho)=\rho \coth{2\rho}-{\rho^2 \over \sinh^2{2\rho}}-{1\over4},\qquad
a(\rho)={2\rho \over \sinh{2\rho}},\qquad
e^{-5 f(\rho)}={2 \sqrt{G(\rho)}\over \sinh{2\rho}}. \label{mnsol}
\end{equation}
In fact as pointed out in \cite{gubser1} the general supersymmetric solution for one scalar and group $SU(2)^+$ is:
\[
G(\rho)=\rho \coth{(2\rho+c)}-\frac{1}{4} a(\rho)^2-{1\over4},\qquad
a(\rho)={2\rho \over \sinh{(2\rho+c)}},\qquad
e^{-5 f(\rho)}={2 \sqrt{G(\rho)}\over \sinh{(2\rho+c)}}.\\[2mm]
\]
The parameter $c$ (which may be assumed positive) is essential: as it changes from 0 to infinity it spans all the BPS solutions, connecting the Abelian one ($c \to \infty$) to the regular, MN one ($c=0$).

Adding the scalar $l(\rho)$ it is still possible to find an
analytic, supersymmetric solution \cite{abcpz}: with the change of
variables $du/d\rho = \cosh{l}$ we get
\begin{equation} l(u)= -\textrm{arcsinh}\left(\frac{c_1}{2 \sqrt{1-8 u^2-\cosh{4 u}+4 u \sinh{4 u}}}\right)\end{equation}
and the same expression  of (\ref{mnsol}) for the other fields with the substitution $\rho \to u$.\\
This background however doesn't flow to the MN solution in the IR and indeed is singular.

Enlarging the gauge group to the other $SU(2)_-$ we get solutions with \N=2 supersymmetry in the UV.
The simpler has only $A_{(1)}$ turned on, that is group $U(1)_+ \times U(1)_-$,
and only two scalars in the coset spaces. It reads \cite{martelli,bcz}:
\begin{equation}
\frac{d u}{d\rho}\equiv e^{l}; \qquad \qquad A= \frac{1}{2} \cos \theta \; d \phi
\end{equation}
\[
G(u)=u; \qquad
l(u)=\frac{1}{2} \log \left[ 1-\frac{1}{2 u}+\frac{2 K e^{-2 u}}{u}\right] ;\qquad
f(u)= \frac{2}{5} G(u) +\frac{1}{5} l(u).\\[3mm]
\]
The free parameter $K$ should satisfy $K \ge \frac{1}{4}$ to have $u \in [0,\infty)$.

A more general case consider the three scalars $f, l, k$ \cite{bcz};
the relations
$d u / d\rho \equiv e^{l}$
and $G(u)=u$ still hold and moreover we have:
\begin{equation}
e^{k(u)}= \frac{e^{2 u}-b^2}{e^{2 u}+b^2}
\end{equation}
We note that all the known \N=2 solutions are singular in the origin, even if it is supposed that the enhan\c con mechanism should regularize the IR behavior. The solutions we will find in the next section are instead regular at the origin by construction.
\subsection{General solutions: the IR behavior}
We require the regularity of the solution, thus for small $u$ we look for expansions of the form
\[ \phi(u) \simeq \phi_0+ \phi_1 \, u+ \phi_2 \, u^2+\ldots \]
for each of the six fields in the lagrangian (\ref{lagmanu}).\\
In the full $SU(2)_+ \times SU(2)_-$ group case, to obtain a simple expression for the UV expansion, is useful to redefine the radial coordinate as $du/dr= exp[-l(u)]$. \\
Solving the second order equations of motion with these ansatz we find:
\begin{equation}
\begin{array}{rccccccc}
a(u)&\simeq&1& +& \a  \, u^2&+&a_4 u^4 &+ \ldots \ , \\
\tilde{a}(u)&\simeq& 1 &+& \b \, u^2&+&\tilde{a}_4 u^4 &+ \ldots \ , \\
l(u)&\simeq&&& \g \, u^2&+&l_4 u^4 &+ \ldots \ , \\
k(u)&\simeq&&&k_2 u^2&+&k_4 u^4&+ \ldots \ , \\
G(u)&\simeq&&& u^2&+&G_4 u^4&+ \ldots \ , \\
f(u)&\simeq&C_f&+&f_2 u^2&+&f_4 u^4&+ \ldots \ .\\
\end{array}
\end{equation}
Here $C_f$ is unfixed, and represents the freedom to shift the constant value of the dilaton.
$\a, \b, \g$ are the free parameters, while the coefficients labeled as $\phi_i$ are determined in function of these. For example
\[
f_2= \frac{2}{15}+\frac{(\a^2+\b^2)}{10},\qquad G_4=-\frac{2}{9}-\frac{\a^2+\b^2}{2},\qquad k_2=\frac{\a \b +\g}{2},\] 
\begin{equation}  a_4=-\frac{2}{9} \a +\frac{3}{10} \a^2-\frac{1}{10} \a^3- \frac{7}{10} \a \b^2 -\b \g, \qquad \ldots .
\end{equation}
\\
For gauge group $SU(2)_+$ only, that is for $\beta=0$ and, for consistency, $k(u)=0$, one can retain the original variable $\rho$ or use the change: $du/dr= \cosh [l(u)]$ with only slight differences. Using the above redefinition we have:
\begin{equation}
\begin{array}{rcccccccccc}
a(u)&\simeq&1&       &+& \a \, u^2&&&+&a_4 u^4 &+ \ldots \ , \\
l(u)&\simeq& &\g \, u&&  &+&l_3 u^3 &&&+ \ldots \ , \\
G(u)&\simeq& &       & &u^2       &&&+&G_4 u^4 &+ \ldots \ ,\\ 
f(u)&\simeq&C_f&     &+ &f_2 u^2   &&&+&f_4 u^4 &+ \ldots \ , 
\end{array}
\end{equation}
with, for example:
\[
l_3=\frac{\g}{90} (-4 +36 \a +27 \a^2 +3 \g^2), \qquad f_2= \frac{2}{15}+\frac{\a^2}{10}, \qquad G_4=-\frac{2}{9}-\frac{ \a^2 + \g^2}{2},\quad \ldots .
\]
and only two free parameters $C_a$ and $C_l$ (plus $C_f$).\\
The MN solution correspond to $C_a=-2/3$, $C_l =0$.
\subsection{General solutions: the complete UV behavior}
We start with the full $SO(4)$ case.

We are searching supersymmetry breaking solutions, so we want that in the far UV the general solution is asymptotic to the original MN one.
Corrections to the supersymmetric background can be of two types, normalizable and non-normalizable, whether we are giving a vev to the corresponding operator or turning on a mass. However, since the supergravity background is not an AdS one, there is no direct correlation between the infinitesimal order of the solution and the dimension of the corresponding operator, such as it is usual in AdS/CFT correspondence and as we will find for the Klebanov Strassler case.
For example all non-normalizable (but still vanishing as $\rho \to \infty$) corrections are polynomial in $1/u$.
Nevertheless it is not difficult to find that corrections are all in the form of a polynomial times a power of $e^{2 u}$.
Thus the expression for the leading terms of UV expansion is:\\[0mm]
\[ G(u) \simeq u+G_{\infty}-\frac{M_a^2+M_{\tilde{a}}^2}{4} \frac{1}{u}+\ldots \\[1mm]\]
\begin{equation} f(u) \simeq \frac{2}{5} u +f_{\infty}+(-\frac{1}{20}+\frac{M_a^2+M_{\tilde{a}}^2}{10}) \frac{1}{u}+\ldots\\[1mm] \end{equation}
\[ a(u) \simeq u^{-\frac{1}{2}} M_a \left[1+\frac{1}{4} (1-2 M_a^2-2 G_{\infty}) \frac{1}{u}+\ldots \right]\\[1mm] \]
\[ \tilde{a}(u) \simeq u^{-\frac{1}{2}} M_{\tilde{a}} \left[1+\frac{1}{4} (1-2 M_{\tilde{a}}^2-2 G_{\infty}) \frac{1}{u}+\ldots \right]\\[1mm] \]
\[ l(u) \simeq \frac{1}{4}\frac{1}{u}+(\frac{1}{16}-\frac{M_a^2+M_{\tilde{a}}^2+G_{\infty}}{4})\frac{1}{u^2} \ldots\\[1mm] \]
\[ k(u) \simeq \frac{M_a M_{\tilde{a}}}{4} \frac{1}{u^2} \left[-1+(-2+2 G_{\infty}+3 \frac{M_a^2+M_{\tilde{a}}^2}{2})\frac{1}{u}+ \ldots \right] \\[2mm]\]
The full expression for the sub-leading expansion is rather involved.
For simplicity in the following formulas we have truncated the dependence of the sub-leading coefficients from the mass parameters $M_i$ to the linear order, which is the only relevant for our purposes. We have also used the freedom to shift $u$ to eliminate the dependence on $G_{\infty}$.\\[2mm]
\[
a(u):e^{-2 u} \sqrt{u} V_a \left[1+\left(\frac{3}{4}+\frac{V_k M_{\tilde{a}}}{V_a}\right) \frac{1}{u}+\left(\frac{7}{32}+\frac{V_k M_{\tilde{a}}}{ 4 V_a} \right) \frac{1}{u^2}+\ldots \right] \\[2mm]
\]
\[\tilde{a}(u):e^{-2 u} \sqrt{u} V_{\tilde{a}} \left[1+\left(\frac{3}{4}+\frac{V_k M_a}{V_{\tilde{a}}}\right) \frac{1}{u}+\left(\frac{7}{32}+\frac{V_k M_a}{ 4 V_{\tilde{a}}} \right) \frac{1}{u^2}+\ldots \right]\\[2mm]
\]
\begin{equation}
G(u): e^{-2 u}  (M_a V_a+M_{\tilde{a}} V_{\tilde{a}}) \left[\left(2 u +\frac{9}{16} \frac{1}{u}+\ldots \right) +V_G \left(1+\frac{5}
{8} \frac{1}{u}
+\ldots\right)\right] \\[2mm]
\end{equation}
\[
l(u):  e^{-2 u} (M_a V_a+M_{\tilde{a}} V_{\tilde{a}}) \left[V_L \left(\frac{1}{u}+\frac{1}{2} \frac{1}{u^2}+\ldots\right)+V_G \left(\frac{1}{4}\frac{1}{u^2}-\frac{1}{8} \frac{1}{u^3}+\ldots\right)+\left(\frac{1}{u^2}+ \ldots\right) \right] \\[2mm]\]
\[f(u): e^{-2 u} (M_a V_a+M_{\tilde{a}} V_{\tilde{a}}) \left[\left( \frac{2}{5}+\frac{3}{10 u}+\ldots \right) -\frac{1}{5} V_L \left(\frac{1}{u}+\frac{1}{2 u^2}+\ldots \right)+V_G \left(-\frac{1}{8 u^2}+\ldots\right) \right] \\[2mm]\]
\[
k(u): e^{-2 u} \left[V_K+ (M_a V_{\tilde{a}}+ M_{\tilde{a}} V_a) \left(\frac{1}{2u}+ \frac{1}{4 u^2}\ldots \right) \right]\\[3mm]
\]
The free parameters $V_G$ and $V_l$ scale in such way that even for the solutions with $M_i \to 0$ the product $(M_a V_a+M_{\tilde{a}} V_{\tilde{a}}) \times V_{G,l}$ remains finite. \\
The \N=1 case, with only $SU(2)_+$ turned on, shows some differences.
It is still possible to find the subleading exponentially suppressed expansion for all the fields:
\begin{eqnarray}
G(u)&:&e^{-2 u} u^{\frac{3}{2}} \left[V_G (1+\frac{1}{2 u}+\frac{1}{8 u^2}+\ldots)-40 V_f (\frac{1}{2 u}+\frac{1}{8 u^2}+\ldots)\right],\nn\\
f(u)&:&e^{-2 u} u^{\frac{1}{2}} \left[-\frac{1}{5}V_G (1+\frac{1}{2 u}+\ldots)+ \frac{V_f}{u^2} (1+\frac{1}{2 u}+\ldots)\right],\\
a(u)&:& e^{-2 u } V_a (u +\frac{1}{4}+\ldots),\nn\\
l(u)&:&e^{-2 u} V_l u^{-\frac{1}{2}}.\nn
\end{eqnarray}
It is not possible however to find a polynomial expansion for all the four fields; when the difference of the scalars is turned on the leading term of the UV expansion becomes probably exponentially divergent.
Indeed, substituting a polynomial expansion for the fields in the equations of motion one gets simply that $l=0$ at all orders, while for $a, G, f$ we have the known expression:
\[a(u)= M_a u^{-\frac{1}{2}} \left[1-\frac{1}{2 u} (-1+M_a^2+G_{\infty})+\ldots\right]\]
\begin{equation}f(u)=\frac{2}{5} u-\frac{1}{10} \log u + f_{\infty} -\frac{G_{\infty}}{10 u}+\ldots \end{equation}
\[G(u)=u + G_{\infty}-\frac{M_a^2}{2 u}+\ldots\]
\subsection{Numerical interpolation}
\begin{figure}
\begin{picture}(400,100)(0,0)
\put(5,0){\epsfig{figure=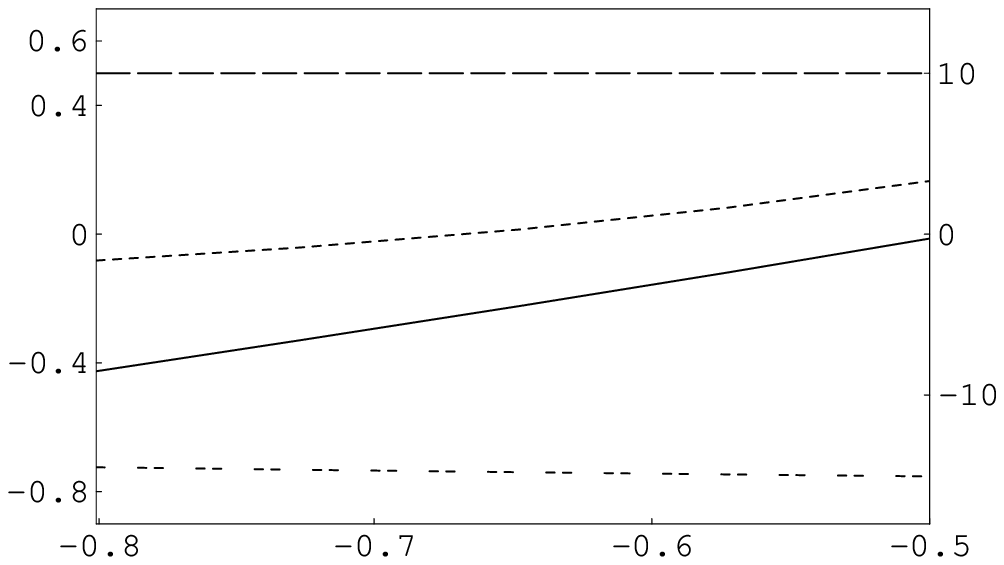,width=7cm}}
\put(240,0){\epsfig{figure=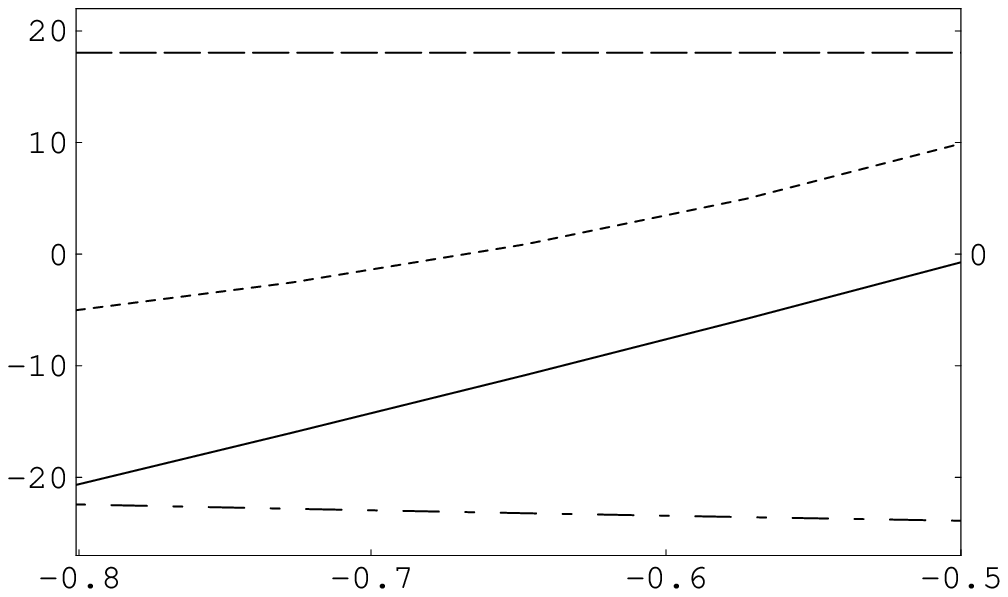,width=7cm}}
\end{picture}
\caption{\footnotesize Dependence of the UV parameters from the IR
parameter of the condensate $a(u)$, when the other IR parameters
vanish. To the left, the dependence of the mass parameters, to the
right that of the vev parameters. The dotted line represents the
behavior of $M_a$ and $V_a$, the solid line that of $G_{\infty}$
and $V_G$, the dash-dotted $f_{\infty}$,$V_f$ and the dashed line
$M_{\tilde{a}}$ and $V_{\tilde{a}}$. The ordinate of $M_a,
f_{\infty}, G_{\infty}$ should be red on the left axes, that of
$M_{\tilde{a}}$ on the right one. The plots for the vev parameters
have been rescaled to fit in the same graphic, the labels of
vertical axes should not be taken as absolute values. Note that
both parameters $M_{\tilde{a}}$ and $V_{\tilde{a}}$ of the second
condensate $\tilde{a}(u)$ do not depend on the first one $a$, as
expected since they are independent operators.}\label{masseinb}
\end{figure}

It is now possible to find globally defined solutions.
First of all the free parameters in UV expansions are not independent from that of IR one. We show for example in fig. \ref{masseinb} the dependence respectively of the four leading parameters and of the four subleading parameters
from the variation of the IR parameter $C_a$, when $\b$ and $\g$ are zero.

\begin{figure}
\begin{picture}(400,230)(0,0)
\put(20,110){\epsfig{figure=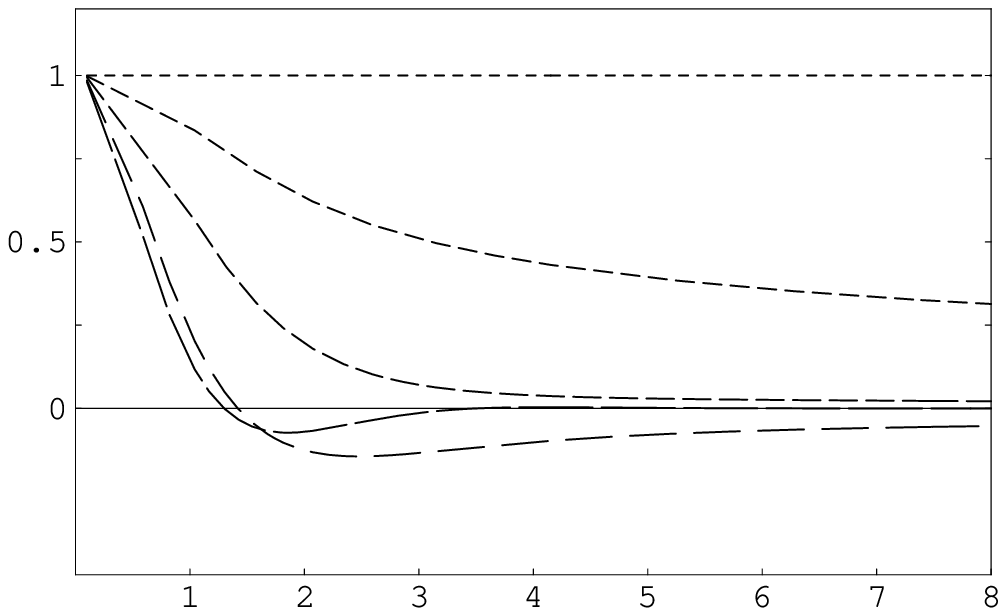,width=6.5cm}}
\put(245,110){\epsfig{figure=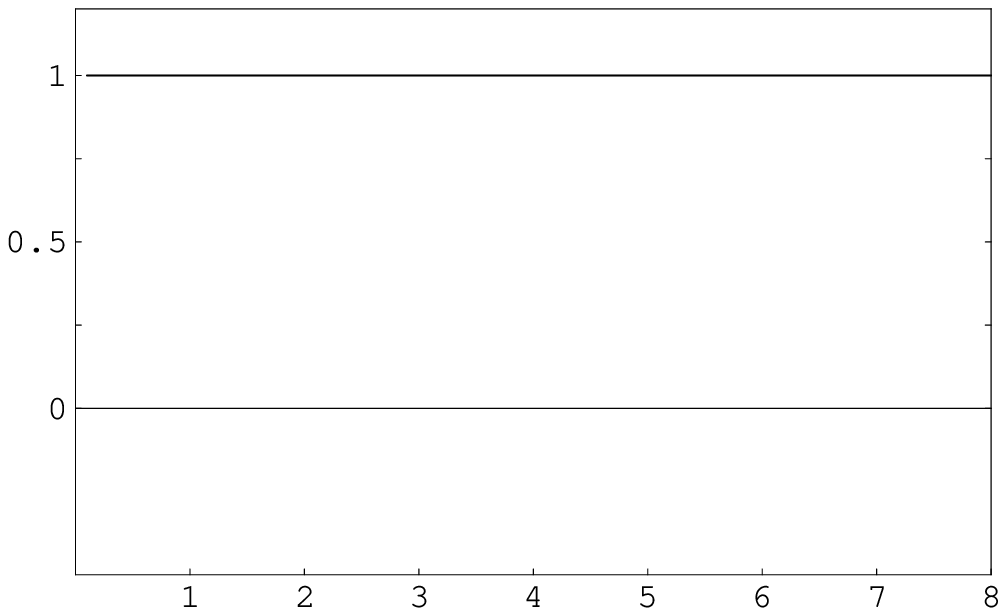,width=6.5cm}}
\put(245,-10){\epsfig{figure=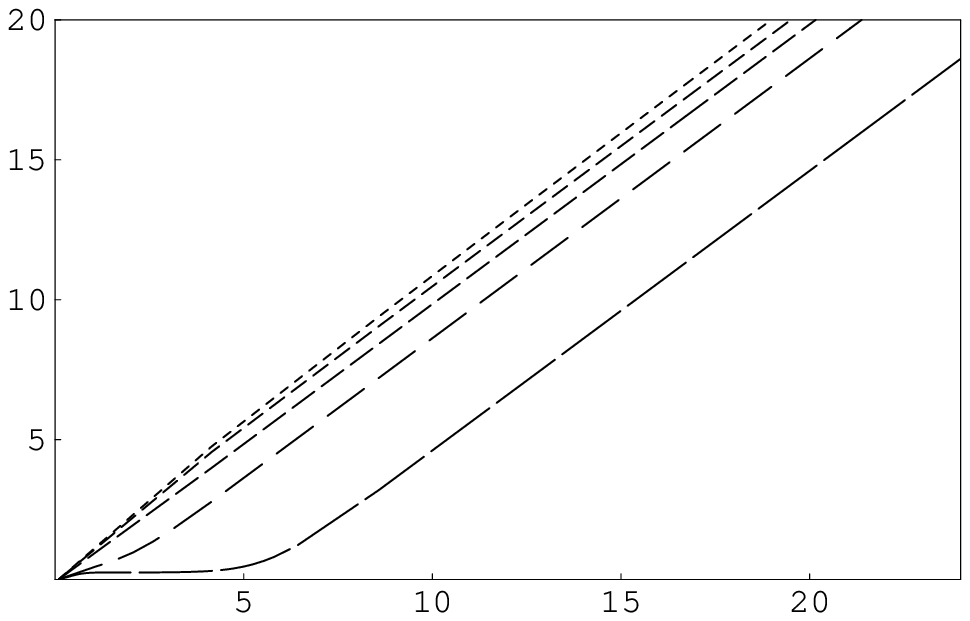,width=6.5cm}}
\put(25,-10){\epsfig{figure=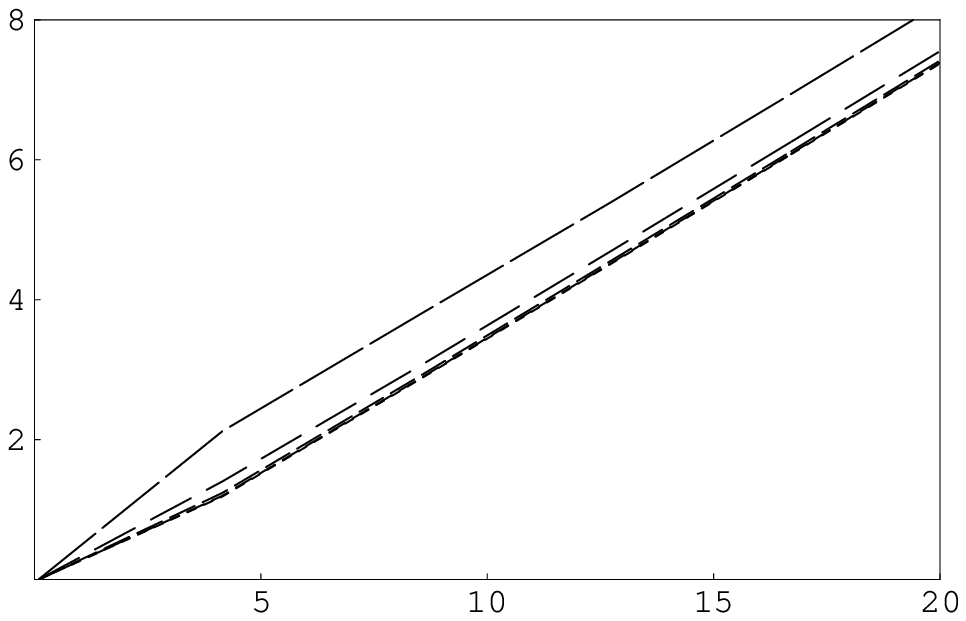,width=6.5cm}}
\put(0,160){\makebox(0,0)[b]{$a(u)$}}
\put(230,160){\makebox(0,0)[b]{$\tilde{a}(u)$}}
\put(230,20){\makebox(0,0)[b]{$f(u)$}}
\put(5,20){\makebox(0,0)[b]{$G(u)$}}
\end{picture}
\caption{\footnotesize Behavior of $SO(4)$ solutions as $C_a$
changes from 0 (shortest-dashed line) to -2 (longest-dashed line),
with the other IR parameters set to zero. For $C_a > 0$ the
solution for the field $a(u)$ diverges. }\label{varia}
\end{figure}

Then, using numerical interpolation, it is possible to find numerical solutions of the equations of motion for all the values of the radius $u$ (or equivalently $r$).
We show for example in fig. \ref{varia} the solutions corresponding to changing $\a$ in the same range of fig. \ref{masseinb}, with $\b$ and $\g$ vanishing.
Note that $a$ vanishes at infinity only if we start with $C_a < 0$. For the limiting case $C_a=0$ we have the $U(1)\times U(1)$ solution with $a(u)=1$.

\begin{figure}
\begin{picture}(400,230)(0,0)
\put(20,110){\epsfig{figure=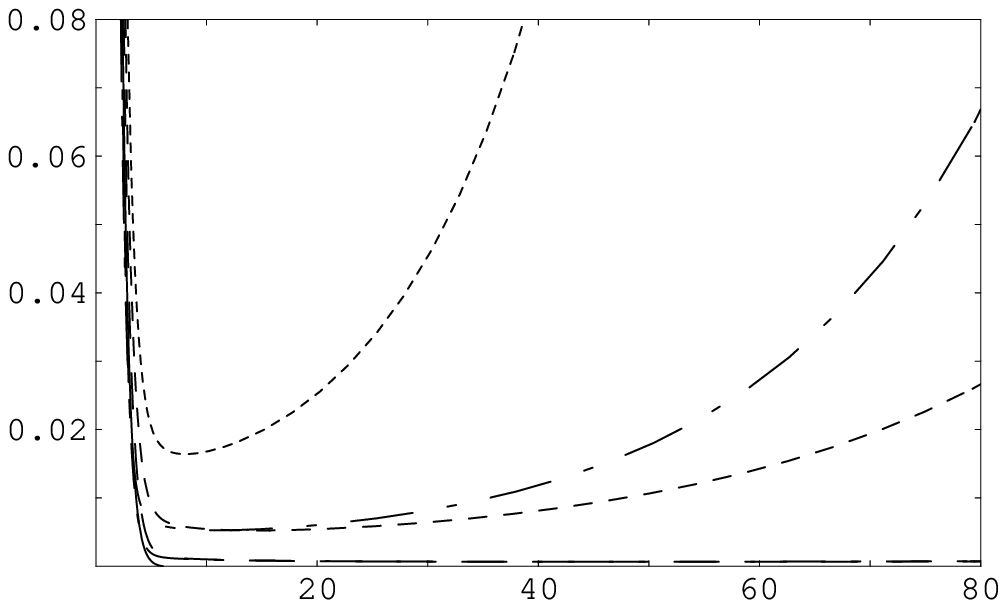,width=6.5cm}}
\put(245,110){\epsfig{figure=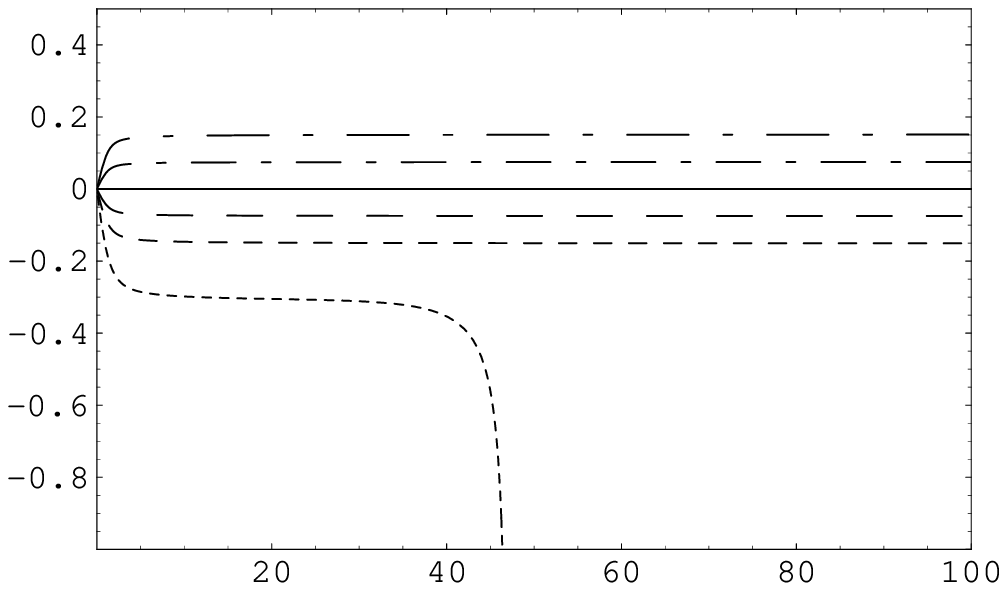,width=6.5cm}}
\put(245,-10){\epsfig{figure=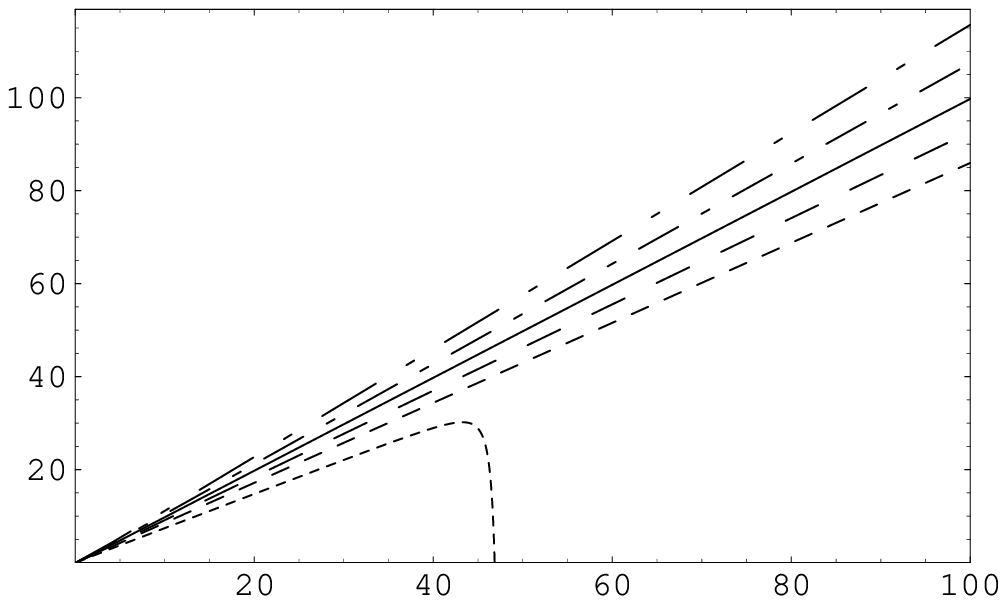,width=6.5cm}}
\put(25,-10){\epsfig{figure=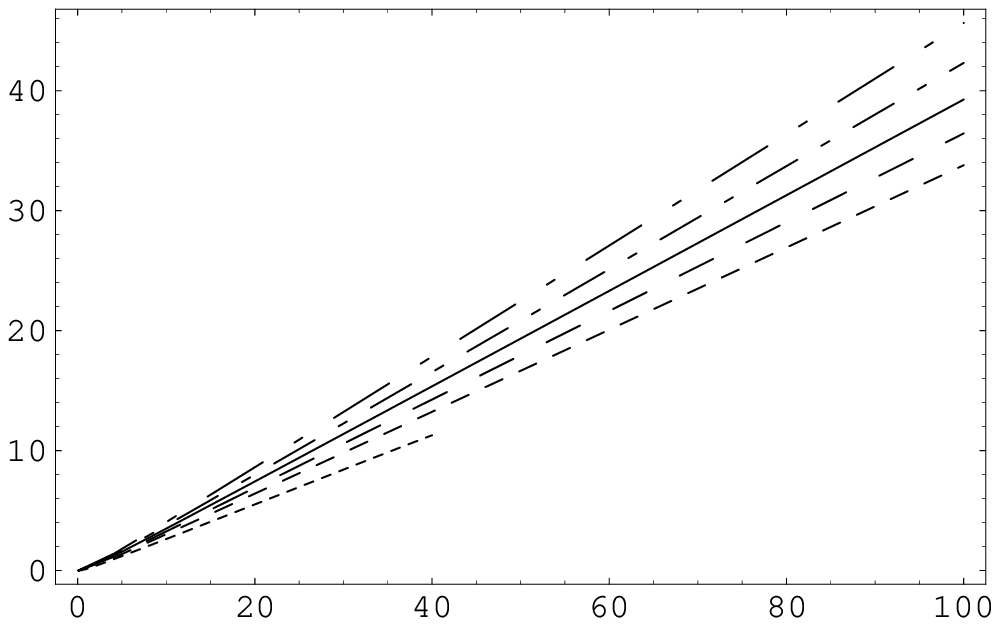,width=6.5cm}}
\put(0,160){\makebox(0,0)[b]{$a(u)$}}
\put(230,160){\makebox(0,0)[b]{$l(u)$}}
\put(230,20){\makebox(0,0)[b]{$f(u)$}}
\put(5,20){\makebox(0,0)[b]{$G(u)$}}
\end{picture}
\caption{\footnotesize Behavior of the solutions for $SU(2)_+$ gauge group for different values of $\g$: $\g=0$ is the solid line, dash-dotted lines have $\g$ positive and increasing while dotted ones have $\g < 0$. Even if not shown in the figure, all solutions with $\g \neq 0$ diverge as $u$ goes to infinity  } \label{accensione2}
\end{figure}

We show in fig \ref{accensione2} what happens to the solution in
the MN setup (gauge group $SU(2)$, $\a=-2/3$) if we turn on
gradually the other free parameter $\g$. We note that all the
numerical solutions we are been able to find for this case do not
connect in the UV to the wanted one with both $a$ and $l$
vanishing. Instead $l(u)$ seems to diverge exponentially for $u$
large enough. As already pointed out, neither the supersymmetric
solution of the previous section flows in the IR to the regular
one. It is not clear whether a globally regular solution exists or
not. 


\subsection{Vacuum energy}
The vacuum energy for various \N=1 deformations of the MN background was computed in \cite{gubser1,epz}. We now generalize the computation to our solutions.
The redefinition:
\[
h(r)=2 g(r)-2 f(r)=\texttt{Log} G(r); \qquad A(r)=3 f(r)+2 g(r)
\]
diagonalizes the kinetic terms of the lagrangians (\ref{lagmanu}),(\ref{lagmanu2}).\\
The result can be  expressed for both cases in the form:
\begin{equation} \mathcal{L}= e^A \left(\dot{A}^2-\frac{1}{2} G_{ab}\dot{\phi}^a \dot{\phi}^b -V(\phi) \right)\end{equation}
where the notation $\phi_a$ indicates the other fields different from $A$.

After integration by parts and the use of the equations of motion,
the action is thus given by
\begin{equation} I \sim \lim_{r \to \infty}  2 \dot{A}(r) e^{A(r)} \end{equation}
For the $N=1$ case this is exactly the expression of \cite{gubser1,epz} and we do not re-discuss the consequent computation here.
For the $N=2$ case, we have to make  the change of variables $dr \to du \; e^{l(u)}$, so that:
\begin{equation} I \sim \lim_{u \to \infty}  2 \dot{A}(u) e^{A(u)-l(u)} \end{equation}
We remark that this expression, even if deduced from the effective uni-dimensional lagrangian, is the same that we
get starting from the full action, in the Einstein frame, and including both the volume and the boundary contributions.

If we substitute the general expression for the fields in the
above formula we have of course a divergent result. Divergences
will cancel however when we compare the energies of any two
solutions, provided that we have correctly matched the two
corresponding metrics on the boundary. We take the BPS zero
gaugino mass spacetime as our reference geometry. For that metric,
using the freedom to shift $u$ and $f_{\infty}$ we find
\[ I \sim \lim_{u \to \infty} \partial_u \left(2 (u+u^*)+\frac{1}{2} \log\left[(u+u^*)^2-\frac{(u+u^*)}{2}\right]+5 (f_{\infty}+ f^*) \right) (u+u^*) e^{2 (u+u^*)+5 (f_{\infty}+f^*)} \]
The matching of the dilaton and of the $S_2$ metric fixes $u^*$ and $f^*$, and corresponds to equating the gauge coupling in the UV.
We have, at linear order in the mass parameters (the label ``S'' indicate the subleading contribution):
\begin{displaymath}
\begin{array}{ccrcl}
5 f(u)-l(u):&\qquad \quad &2 (u+u^*)+5(f_{\infty}+f^*) &\quad \equiv \quad& 2 u+5 f_{\infty}+5 f_S (u)-l_S (u)  \\
G(u):& \qquad \quad &u+G_{\infty}+u^* &\quad \equiv \quad& u+G_{\infty}+G_S (u)
\end{array}
\end{displaymath}
Thus
\begin{equation} u^*\sim G_S (u)+\ldots \qquad \quad 5 f^* \sim 5 f_S(u)-l_S(u)-2 G_S(u)+\ldots \end{equation}
When comparing two vacuum energies this leads to
\begin{equation}
\Delta I \sim \lim_{u \to \infty} u e^{2 u+5 f_{\infty}}
\left[\dot{l_S}+\frac{1}{4 u^2} \dot{G_S}+O
\left(\frac{G_S}{u^3}\right)\right]
\end{equation}
Substituting the expressions for the subleading terms we finally have:
\begin{equation}
\Delta I \propto e^{2 \Phi_{\infty}}(V_l+1) (Re (M_a V_a) +Re (M_{\tilde{a}} V_{\tilde{a}}))
\end{equation}
For simplicity we have not shown the contributions of high order in the masses
in the intermediate steps. However in the difference between solutions all contributions that are not vanishing by
themselves in the limit $u \to \infty$ exactly cancel, and the final result for $\Delta I$ does not change.

It is clear that in the supersymmetric limit when the masses of the two gauginos are both vanishing the solutions
with different phases for the two condensates are degenerate. When we introduce instead a soft supersymmetry-breaking
mass term for one gaugino (or both) this degeneracy is lifted by a term proportional to the mass itself
(or to the sum of the masses), as expected from gauge theory analysis.
\section{Gravity duals of gauge theories from regular and fractional D3 branes on a deformed conifold}
\subsection{Branes setup}
Putting branes on singular backgrounds is a promising way of
breaking supersymmetries in the context of gauge
theory/supergravity correspondence. In particular it is possible
to have a dual of a $N=1$ theory putting N D3-branes at the tip of
a singular Calabi-Yau space known as the conifold. There are
various way to identify this space: it is the complex manifold
described by the equation $\sum_i z_i^2 =0$; it has metric
\[
d s^2_6= d r^2+ r^2 d s^2_{T^{1,1}}= d r^2+r^2 \frac{1}{9}( d \psi
+ \sum_{i=1}^2 \cos \theta_i \phi_i)^2+\frac{1}{6}\sum_{i=1}^2( d
\theta_i^2+ \sin^2 \theta_i d \phi^2_i);
\]
the base of the conifold $T^{1,1}$ is the coset space $(SU(2)\times SU(2))/U(1)$ with topology $S^2 \times S^3$.

The spacetime we get for the system of N ordinary D3 branes placed at the origin of the conifold in the supergravity limit is $AdS_5 \times T^{1,1}$. The dual gauge theory is conformal, has $N=1$ SUSY and gauge group $SU(N)\times SU(N)$ coupled to chiral superfields (two bifundamentals $A_1$, $A_2$ transforming in the $(N,\bar{N})$ representation and their two conjugates $B_1$, $B_2$ transforming in the $(\bar{N},N)$) \cite{klewi}.
At this level the supergravity background remain singular.

Adding M fractional D3 branes, which in this context can be viewed as D5 branes wrapped on a collapsing two cycle of the $T^{1,1}$, changes the gauge group to $SU(N)\times SU(N+M)$ and breaks the conformal symmetry \cite{KS}. The relative gauge coupling starts to run logarithmically, and so does the D3 brane charge. Along the RG flow the theory experiences a series of Seiberg dualities, in which the gauge group factors are repeatedly lowered by M units, until finally in the far IR the gauge group become simply $SU(M)$ \footnote{Anyway even for the KS case we do not get a pure SYM, since for $g_s M \gg 1$, where supergravity is a valid approximation, the duality cascade is dense. There is no finite energy range in which extra modes decouple.}. Non perturbative effects become essential and the exact $Z_{2 M}$ R-symmetry is dynamically broken to $Z_2$.
The chiral symmetry breaking of gauge theory corresponds in supergravity to the deformation of the conifold, which resolves the singularity. This deformation is obtained keeping the $S_3$ of $T^{1,1}$ of finite size even in the origin, so that the D5 branes no longer concentrate but smear over the $S^3$, keeping the  3-form flux finite.
The deformed conifold is now described by the equation $\sum_i z_i^2=\epsilon^2,$ with $\epsilon$ given by the mass scale of the gauge theory: $\epsilon^2 \sim m^3$. The supergravity background becomes a sort of warped $AdS_5$ times the deformed conifold base. 
\subsection{Supergravity action and supersymmetric solution}
We start from the usual ten dimensional type IIB action:
\begin{eqnarray}  S_{ 10}
= - \frac{1}{2 \k_{10}^2}  \int d^{10} x \bigg( \sqrt{-g_{10}} \big[ \ R_{10}
 -  \texttt{ $\frac{1}{2}$} (\p \P)^2
- \texttt{$\frac{1}{12}$} e^{-\P}   (\p B_2)^2
 \\  -  \quad \texttt{$\frac{1}{2}$}  e^{2 \P} (\p \C)^2
 - \texttt{$\frac{1}{12}$} e^{ \P}  (\p C_2  - \C  \p B_2) ^2
- \texttt{$\frac{1}{4\cdot 5!}$}  F^2_5\ \big]
- \texttt{$\frac{1}{2\cdot 4! \cdot (3!)^2 }$}
 {\ep_{10}} C_4 \p C_2 \p B_2 + ... \bigg) \ ,  \nonumber \\[3mm]
(\p B_2)_{MNK} \equiv 3 \p_{[M} B_{NK]}\ , \ \
\  \  \ (\p C_4)_{MNKLP} \equiv 5 \p_{[M} C_{NKLP]} \ , \nonumber \\
 F_5= \p C_4 + {5} (B_2 \p C_2 - C_2 \p B_2) \  , \nonumber
\end{eqnarray}
with the additional on-shell constraint $F_5 = *F_5$.\\
To describe the brane setup of the model we take the following ansatz for the metric:
\begin{equation}
ds^2 = e^{2 p - x}(dr^2+e^{2A(r)}dx_{\mu}dx^{\mu})+e^{-6 p -x} g_5^2+e^{x+y}(g_1^2+g_2^2)+
e^{x-y}(g_3^2+g_4^2)
\end{equation}
This is the general form for a warped and deformed version of the original $AdS_5 \times T^{1,1}$. The 1-forms $g_i$ are combinations of the angular differential 1-forms:
\pagebreak
\[
g_1 =\frac{ -\sin \theta_1 \phi_1- \cos \psi \sin \theta_2 d \phi_2 + \sin \psi d \theta_2}{\sqrt{2}}, \qquad 
g_2 =\frac{ d \theta_1 - \sin \psi \sin \theta_2 d \phi_2 - \cos \psi d \theta_2}{\sqrt{2}},
\]
\[
g_3= \frac{-\sin \theta_1 \phi_1+ \cos \psi \sin \theta_2 d \phi_2 - \sin \psi d \theta_2}{\sqrt{2}}, \qquad 
g_4= \frac{ d \theta_1 + \sin \psi \sin \theta_2 d \phi_2 + \cos \psi d \theta_2}{\sqrt{2}},
\]
\[
g_5=d \psi +\cos \theta_1 d \phi_1+ \cos \theta_2 d \phi_2
\]
The function $y$ describes the asymmetry between the two $S_2$.\\
In order to diagonalize the kinetic terms of the lagrangian preserving the same normalization of \cite{KleTs}, we introduce the fields $q$ and $f$:
\begin{equation}
p= -q+f +\frac{1}{4} \log 3 \, , \qquad \qquad x = 3 q + 2 f -\frac{1}{4} \log 12.
\end{equation}
We can safely take $C=0$, while the expressions for the other forms are constrained by the closure conditions and by the symmetries of the system.
Finally, to avoid confusion with future notations, we rename the number $M$ of fractional D-branes as $P$, and the number $N$ of regular branes as $Q$.
So the ansatz for all the fields becomes:
\begin{eqnarray}
ds^2&=&\sqrt{2}\: 3^{\frac{3}{4}} \left( e^{-5q(r)}(dr^2+e^{2A(r)}dx_{\mu}dx^{\mu})
+ds^2_{5\prime}\right)\nonumber\\
ds^2_{5\prime}&=& \texttt{$\frac{1}{9}$} e^{3q-8f}g_5^2+\texttt{$\frac{1}{6}$} e^{3q+2f+y}(g_1^2+g_2^2)+ \texttt{$\frac{1}{6}$}
e^{3q+2f-y}(g_3^2+g_4^2)\nonumber\\
\C &=& \C_4 = 0 \, ,\nn \\
B_{(2)}&=&g(r)\ g_1\wedge g_2+k(r)\  g_3\wedge g_4\nonumber\\
\p \C_2 \equiv F_{(3)}&=&2 P g_5\wedge g_3\wedge g_4+d[F(r)(g_1\wedge g_3+
g_2\wedge g_4)]\nonumber\\
F_{(5)}&=& L(r)g_1\wedge g_2\wedge g_3\wedge g_4\wedge g_5
+ *(L(r)g_1\wedge g_2\wedge g_3\wedge g_4\wedge g_5)\nn \\
L(r)&=& Q+ (k(r)-g(r)) F(r) + 2 P g(r) .
\end{eqnarray}
It is now easy to derive an effective action for the relevant fields \cite{KleTs,PaZ}:
\begin{eqnarray}
L&=&\int dr e^{4A}\left ( 3(\partial A)^2-{1\over 2}G_{ab}\partial\phi^a
\partial\phi^b -V(\phi )\right )\nonumber\\
&&\nn \\
G_{ab}\partial\phi^a\phi^b&=&15(\partial q)^2+10(\partial f)^2+ {(\partial y)^2\over 2}+
{(\partial\Phi)^2\over 4}\nonumber\\
 &+&
 e^{-\Phi-6q-4f}
(e^{-2y}{\sqrt{3}(\partial g)^2\over 2}+e^{2y}{\sqrt{3}(\partial k)^2\over 2})+
 \sqrt{3} e^{\Phi
-6q-4f}{(\partial F)^2}\nonumber\\
V(\phi)&=& e^{-8q}\left (e^{-12f}-6e^{-2f}\cosh{y}+{9\over 4}e^{8f}
\sinh{y}^2\right )+ {9\sqrt{3}\over 8}e^{4f-14q}
e^{-\Phi}(g-k)^2\nonumber\\
&+& {9\sqrt{3}\over 4}e^{4f-14q+\Phi}
(e^{-2y}F^2+e^{2y}(2 P - F)^2)
+{27\over 2}e^{-20q}
L^2.
\label{potks}
\end{eqnarray}
Einstein equations also impose the
constraint $3(\partial A)^2-{1\over 2}G_{ab}\partial\phi^a
\phi^b +V(\phi )=0$.\\
The undeformed conifold case is obtained for $y=0,\ k=g,\ F=0$.\\
It is useful now to introduce the new radial coordinate $\t$ so that
\begin{equation} d \t \equiv e^{4 p} d r\end{equation}
The supersymmetric solution of \cite{KS} reads
\[
g(\t)=\frac{\t \coth \t -1}{2 \sinh \t}(\cosh \t-1),\qquad \qquad
k(\t)=\frac{\t \coth \t -1}{2 \sinh \t}(\cosh \t +1), \qquad \]
\begin{equation}
F(\t)=\frac{\sinh \t -\t}{2 \sinh \t},\qquad \qquad
e^y= \tanh \frac{\t}{2}, \qquad \qquad \Phi=0,\qquad \\[2mm]
\end{equation}
\[
\qquad e^{10 f}= K \sinh \t,
\qquad e^{15 q}= h^{5/2} K^2 (\sinh \t)^4,
\qquad e^{2 A}= h^{1/3} K^{2/3} (\sinh \t)^{4/3},\qquad \qquad
\]
\[
K(\t)=\frac{(\sinh (2 \t)- 2 \t )^{1/3}}{2^{1/3} \sinh \t} \qquad \qquad
h(\t)\propto \int_\t^{\infty} d x \frac{x \coth x -1}{\sinh^2 x} 
(\sinh (2 x)- 2 x)^{1/3} \qquad \\[2mm]\]
\subsection{Gauge theory: identification of the fields}\label{glui}
The supergravity background for the KS model is not a regular $AdS$ one. Nevertheless, since it is only a deformation of an $AdS$ spacetime, we can guess that the $AdS$ prescription that links the masses of the supergravity fields to their asymptotic UV behaviors and to the dimensions of the corresponding operators of field theory, still holds.
According with this prescription a field of mass $m$ is dual to an operator
of dimension $\Delta$ given, for all the cases we are going to consider in this paper, by \\[-2mm]
\[
\Delta=2+ \sqrt{4+m^2}.\\[-0mm]
\] 
The two independent solutions for that field behave in the UV as the powers $\Delta$ and $4-\Delta$ of the exponential of minus the radius.

 Looking at the lagrangian (\ref{potks}) we can extract the masses for the various fields in the undeformed case. First we have to make the redefinitions $s=k+g$, $d=k-g$, $N_1=F+d/2$, $N_2=d/2-F$ in order to diagonalize the mass matrix. We get that $A, \ s, \ \Phi$ are massless, $m^2_q=32$, $m^2_f=12$, $m^2_y=-3$, $m_{N_1}=21$, $m_{N_2}=-3$.

According with the above $AdS$ prescription we thus expect that the two independent solutions for each field behave as the following power of $e^{-\t/3}$ (we remember that, due to the different conifold definition for the metric, this is the right quantity in comparison with the usual $AdS$ one), each multiplied by a certain polynomial of $\t$ to account for the deformation of the background:
\\[-1.5mm]
\[
\Phi,\ A, \  s: 4,0; \qquad y, \ N_2: 3,1; \qquad f: 6,-2; \qquad N_1: 7,-3;
\qquad q: 8,-4;\\[1mm]
\]
We will see in the next section that the explicit asymptotic solutions derived from the equations of motion have really these features.
Since the UV behaviors confirm the reliability of $AdS$ prescriptions even for the warped $AdS$ space of this model, we can link the masses of the supergravity fields with the dimensions of the corresponding gauge theory operators.

The field $s=k+g$ does not appear in the potential of (\ref{potks}): it is the massless field $\int_{S_2} B_{(2)}$ associated with a marginal direction in the CFT; the corresponding operator is $\frac{1}{g^2} (F_{(1)}+F_{(2)})$ \cite{klewi}.
For the other fields we have $\Delta_q=8$, $\Delta_{N_1}=7$, $\Delta_f=6$, $\Delta_y=3$, $\Delta_{N_2}=3$.
Thus, following the notations and the results of \cite{bgz,cddf}, we can identify these 
fields with operators in the multiplets:
\begin{eqnarray}
&&\nn \\[-8mm]
q,f &\rightarrow& \Tr(W^2 \bar{W}^2) \nn \\
N_1 &\rightarrow& \Tr(A\bar{A}+ B \bar{B}) W^2 \nn \\
N_2 &\rightarrow& \Tr (W^2_{(1)}+ W^2_{(2)})  \\
y &\rightarrow& \Tr (W^2_{(1)}- W^2_{(2)})\nn
\end{eqnarray}
In particular $N_2$ and $y$ contain the two operators associated to the condensates. Indeed, in pure SYM (that is in the far IR) the gaugino condensate is a protected operator of dimension 3, and it was argued in \cite{KS} that also in the UV, for $N \gg M$, the anomalous dimensions of such operators are only of order $O(M/N)$ or less.

We define as tr$\lambda \lambda$ the combination of the two condensates which is charged under the broken $U(1)_R$; it has to be described by a supergravity object that transforms by a phase when $\psi \to \psi + \delta \psi$. 
In \cite{sonlow} this combination was thus associated with one polarization of $C_2=C_2^{RR}+ i B_2^{NS}$. We see that both reasonings lead to the same result. In \cite{sonlow} the authors also show that for KS solution the above identification for the condensate is in agreement with field theory. To do the same for our class of solutions we need the UV behavior of the supergravity fields, thus we postpone the question after the next section.
\subsection{General solution: asymptotic behavior in the IR}
We now look for solutions of the full second order equations of motion.
Requiring that fields must approach the KS background in the IR, we found:
\setlength\arraycolsep{2pt}
\begin{equation}
\begin{array}{cccclccccccccccc}
\\[-4mm]
\texttt{\large{$A$}}_{\rm IR}(\t) &\simeq&\; \frac{2}{3}\; \texttt{Log}(\t) &+& \frac{1}{6}\texttt{Log}(\frac{A_0}{32})&&&+&a_2 \t^2&&&+&a_4 \t^4&&&+ \; \;\ldots \\[2mm]
\texttt{\large{$q$}}_{\rm IR}(\t) &\simeq&\frac{4}{15} \; \texttt{Log}(\t) &+& \frac{1}{6}\texttt{Log}(A_0 \, 3^{\frac{9}{10}} \, 2^{-\frac{7}{5}})&&&+&q_2 \t^2&&&+&q_4 \t^4&&&+ \; \;\ldots \\[2mm]
\texttt{\large{$f$}}_{\rm IR}(\t) &\simeq&\frac{1}{10} \; \texttt{Log}(\t) &+& \frac{1}{10}\texttt{Log}(\frac{2}{3})&&&+&f_2 \t^2&&&+&f_4 \t^4&&&+ \; \;\ldots \\[2mm]
\texttt{\large{$y$}}_{\rm IR}(\t) &\simeq&\quad \; \texttt{Log}(\t) &-& \texttt{Log}(2)&&&+&C_y\, t^2&&&+&y_4 \t^4&&&+ \; \;\ldots\\[2mm]
\texttt{\large{$\Phi$}}_{\rm IR}(\t) &\simeq&&&\Phi_0&&&+&\Phi_2 \t^2&&&+&\Phi_4 \t^4&&&+ \; \;\ldots\\[2mm]
\texttt{\large{$F$}}_{\rm IR}(\t) &\simeq&&&&&&&C_F \, \t^2&&&+&\texttt{\small{$F$}}_4 \t^4&&&+ \; \;\ldots \\[2mm]
\texttt{\large{$k$}}_{\rm IR}(\t) &\simeq&&-&\frac{Q}{2\, P}&+&C_k \t&&&+&k_3 \t^3&&&+&k_5 \t^5&+ \; \;\ldots \\[2mm]
\texttt{\large{$g$}}_{\rm IR}(\t) &\simeq&&-&\frac{Q}{2\, P}&&&&&+&C_g \t^3&&&+&g_5 \t^5&+ \; \;\ldots\\[3mm]
\end{array}
\end{equation}
where $A_0,\Phi_0,C_y,C_F,C_g,$ and $C_k$ are free parameters, while the other coefficients are determined in function of these.
For example:
\begin{eqnarray}
f_2 &=&\frac{-7 A_0 - 52 A_0 C_y+ e^{\Phi_0} ( 20 P^2+80 C_F^2)+ e^{-\Phi_0}(720 C_g^2-5 C_k^2)}{200 A_0}\nn \\[1mm]
a_2&=&
\frac{18 A_0+48 A_0 C_y- 40 e^{\Phi_0} P^2+ e^{-\Phi_0}(720 C_g^2-15 C_k^2)}{180 A_0}
\\[1mm]
q_2&=&\frac{72 \, A_0+192 \, A_0 \, C_y-e^{\Phi_0} (720 C_F^2+220 P^2)+e^{-\Phi_0} (720 C_g^2-105 C_k^2)}{1800 A_0} \nn
\end{eqnarray}
In fact $A_0$ is not a relevant parameter, since it is related to the freedom to shift $\t$ to $\t+\t^*$ in the Lagrangian;
in the same way, one of the three parameters $C_F,C_g,C_k$ only represents the freedom to shift the zero value of $s(\t)=g(\t)+k(\t)$, which is used in \cite{KS} to eliminate the dependence on $Q$ in the solutions. 
Thus we have 4 physical free parameters, corresponding to the two gauge couplings and to the two masses, as expected from field theory analysis.

\subsection{General solution: asymptotic behavior in the UV and interpolation}
Solving the equations of motion in the UV we find that all fields can be expanded in powers of $e^{\t/3}$, each power multiplied by a certain polynomial.
In this expansion all coefficients are determined in terms of a few number of free parameters only. These free parameters characterize the independent solutions, the ones that exist even if the other fields are switched off. Moreover, they appear in the polynomial multiplying exactly the power of $e^{\t/3}$ we would have expected from the $AdS$ prescription.

As concern the two independent solutions of each field, they have different features. Some fields start with exponentially divergent non renormalizable solution, while for other fields this non renormalizable solution is only polynomial. Finally, the remaining fields have both the independent solutions exponentially suppressed.
Of course each field, which would have presented in the UV expansion two exponential orders as the independent solutions, actually  gets non independent, induced contributions at other orders of $e^{\t/3}$ when the other fields are switched on. 

For sake of simplicity we do not show the exponentially divergent solutions, which are not asymptotic to the KS one.
The relevant terms in the UV expansion for the eight fields are thus, for the polynomial part: 
\begin{equation}
\begin{array}{ccl}
A(\t)&\simeq&\frac{1}{3} \t+\frac{1}{6} \texttt{Log}(\t-\frac{1}{4})+A_{\infty}+a_{0,{\scriptscriptstyle -} 1} \frac{1}{\t}+a_{0, {\scriptscriptstyle -} 2} \frac{1}{\t^2}+ \ldots +O(e^{-\frac{2}{3} \t})\\[2mm]
q(\t)&\simeq&\frac{1}{6} \texttt{Log}(\t-\frac{1}{4})+\frac{1}{6} \texttt{Log}(3 \sqrt{3} e^{\Phi_{\infty}} P^2)+q_{0,{\scriptscriptstyle -}1} \frac{1}{\t}+q_{0,{\scriptscriptstyle -}2} \frac{1}{\t^2}+ \ldots + O(e^{-\frac{2}{3} \t})\\[2mm]
f(\t)&\simeq& O(e^{-\frac{2}{3} \t})\\[2mm]
y(\t)&\simeq& O(e^{-\frac{1}{3} \t})\\[2mm]
\Phi(\t)&\simeq&\Phi_{\infty}+O(e^{-\frac{2}{3} \t})\\[2mm]
g(\t)&\simeq& e^{\Phi_{\infty}} P \t+ S_{\infty}+ O(e^{-\frac{1}{3} \t})\\[2mm]
k(\t)&\simeq&  e^{\Phi_{\infty}} P \t+ S_{\infty}+ O(e^{-\frac{1}{3} \t})\\[2mm]
F(\t)&\simeq&  P + O(e^{-\frac{1}{3} \t})
\end{array}\label{UVKSpol}
\end{equation}
\begin{eqnarray}
a_{0,{\scriptscriptstyle -}1} &=&  q_{0,{\scriptscriptstyle -}1} \; = \; \frac{Q+2 e^{\Phi_{\infty}} P^2+ 2 S_{\infty} P}{12 e^{\Phi_{\infty}} P^2}\, , \nn\\
a_{0,{\scriptscriptstyle -}2} &=& q_{0,{\scriptscriptstyle -}2} \; = \;\frac{(Q+2 e^{\Phi_{\infty}} P^2+ 2 S_{\infty} P) \, (Q+ e^{\Phi_{\infty}} P^2+ 2 S_{\infty} P)}{48 e^{2 \, \Phi_{\infty}} P^2}\, , \qquad \ldots \nn
\end{eqnarray}
This is the general expression one gets from the equations of motion.
As in the IR there is in fact the freedom to shift the constant values of $A,$ $\Phi$ and $s=g+k$.
For the KS solution this freedom is used to set $Q=0$, $S_{\infty}=-P$, $\Phi_{\infty}=0$ and $A_{\infty}=\frac{1}{6} \texttt{Log}(2^{-6} 3 e^{\Phi_{\infty}} P^2)$.
For the general solution it is more useful to keep all
this quantities unfixed. For sake of clearness however in the following we leave only
$\Phi_{\infty}$ unfixed and show the simpler coefficients that one gets using  redefinition freedom to eliminate the dependence on Q and to set $S_{\infty}=- P e^{\Phi_{\infty}}$.\\
For the exponentially suppressed part of the expansions we have:
\[
\begin{array}{ccl}
g(\t) &: \quad&-M_2 e^{-\frac{1}{3} \t}+e^{-\frac{2}{3} \t} (g_{2,2} \t^2+g_{2,1} \t  +g_{2,0})+e^{- \t} (g_{3,5} \t^5+ \ldots+g_{3,1} \t- V_2)\\&&+e^{-\frac{4}{3} \t} (g_{4,6} \t^6+\ldots+g_{4,1} \t+V_s)+e^{-\frac{5}{3} \t} (g_{5,7} \t^7+\ldots)+e^{-2 \t} (g_{6,9} \t^9+\ldots)\\&&+e^{-\frac{7}{3} \t} (g_{7,11} \t^{11}+\ldots)+e^{-\frac{8}{3} \t} (g_{8,12} \t^{12}+\ldots)+\ldots\\
&&\\
k(\t)&: \quad& M_2 e^{-\frac{1}{3} \t}+e^{-\frac{2}{3} \t} (k_{2,2} \t^2+k_{2,1} \t  +k_{2,0})+e^{- \t} (k_{3,5} \t^5+ \ldots+ k_{3,1} \t+ V_2)\\&&+e^{-\frac{4}{3} \t} (k_{4,6} \t^6+\ldots+k_{4,1} \t+V_s)+e^{-\frac{5}{3} \t} (k_{5,7} \t^7+\ldots)+e^{-2 \t} (k_{6,9} \t^9+\ldots)\\&&+e^{-\frac{7}{3} \t} (k_{7,11} \t^{11}+\ldots)+e^{-\frac{8}{3} \t} (k_{8,12} \t^{12}+\ldots)+\ldots
\\
&&\\
y(\t)&: \quad&(M_1 \t+y_{1,0}) e^{- \frac{1}{3} \t}+ e^{- \t} (y_{3,4} \t^4+\ldots+y_{3,1} \t +V_1)+e^{-\frac{5}{3} \t} (y_{5,7} \t^7+\ldots)\\&&+e^{-\frac{7}{3} \t} (y_{7,10} \t^{10}+\ldots)+\ldots
\\
&&\\
F(\t)&: \quad&(F_{1,1} \t +F_{1,0}) e^{- \frac{1}{3} \t}+e^{- \t} (F_{3,5} \t^5+ \ldots+P V_1 \t+ F_{3,0})
+ e^{-\frac{5}{3} \t} (F_{5,7} \t^{7}+\ldots)\\&&+e^{-\frac{7}{3} \t} (F_{7,11} \t^{11}+\ldots+V_{N_1} \t+F_{7,0})+\ldots
\\
&&\\
\Phi(\t) &: \quad& e^{-\frac{2}{3} \t} (\Phi_{2,1} \t  +\Phi_{2,0})+e^{-\frac{4}{3} \t} (\Phi_{4,5} \t^5+\ldots +V_{\Phi} t+\Phi_{4,0})+ e^{-2 \t} (\Phi_{6,8} \t^8  +\ldots)\\&&+ e^{-\frac{8}{3} \t} (\Phi_{8,11} \t^{11}  +\ldots)+\ldots
\end{array}
\]
\begin{eqnarray}
A(\t)&: \quad& e^{-\frac{2}{3} \t} (A_{2,2} \t^2+\ldots  +A_{2,-1} \frac{1}{\t}+\ldots)+e^{-\frac{4}{3} \t} (A_{4,5} \t^5+\ldots )+e^{-2 \t} (A_{6,9} \t^9+\ldots )\nn \\&&+e^{-\frac{8}{3} \t} (A_{8,12} \t^{12}+\ldots )+\ldots
\nn\\
&&\nn\\
q(\t)&: \quad& e^{-\frac{2}{3} \t} (q_{2,2} \t^2+\ldots  +q_{2,-1} \frac{1}{\t}+\ldots)+e^{-\frac{4}{3} \t} (q_{4,5} \t^5+\ldots )+e^{-2 \t} (q_{6,9} \t^9+\ldots)\nn\\&&+e^{-\frac{8}{3} \t} (q_{8,12} \t^{12}+\ldots+V_q \t+ \ldots)+\ldots
\nn \\
&&\nn \\
f(\t)&: \quad& e^{-\frac{2}{3} \t} (f_{2,2} \t^2+f_{2,1} \t  +f_{2,0})+e^{-\frac{4}{3} \t} (f_{4,5} \t^5+ \ldots+f_{4,0})\nn\\&&+e^{-2 \t} (f_{6,9} \t^9+\ldots+f_{6,1} \t  +V_f)+e^{-\frac{8}{3} \t} (f_{8,12} \t^{12}+\ldots)+\ldots\label{UVKSexp}\\[-2mm]&&\nn
\end{eqnarray}
where we have labeled the free parameters as $M_a$, $V_a$ according whether they correspond to a mass or to a vev (actually there is a certain ambiguity in choosing which combinations of the coefficients $M_{1,2}$ and $V_{1,2}$ correspond exactly to the masses and vevs of the two gauginos)
, and the derived coefficients as $\phi_{i,j}$, with $i$ order of the negative exponential and $j$ of the power of $\t$ multiplying the coefficient itself.

The derived coefficients that will be useful in future computations are, at linear order in the mass parameters: 
\[
y_{1,0}= -\frac{4}{3 P} e^{\Phi_{\infty}}M_2+2 M_1 ,\qquad
F_{1,1}= \frac{3}{2} P M_1, \qquad F_{1,0}=-3 e^{-\Phi_{\infty}} M_2+ \frac{33}{8} P M_1,\\[1mm]
\]
\[
F_{3,1}=P V_1+O(M_i^3), \quad F_{3,0}= P V_1-e^{- \Phi_{\infty}} V_2+O(M_i^3),\quad
k_{3,1}=-g_{3,1}=e^{\Phi_{\infty}} P V_1+O(M_i^3), \\[2mm]
\]
\[
A_{4,1}=-\frac{1}{2} M_1 V_1 + O(M_i^4), \qquad A_{4,0}=-\frac{1}{8}V_{\Phi}-\frac{13}{16} M_1 V_1+\frac{5}{6 P}e^{-\Phi_{\infty}} M_2 V_1 + O(M_i^4),\\[2mm]
\]
\[
q_{4,1}=-\frac{1}{20} M_1 V_1 + O(M_i^4), \qquad q_{4,0}=-\frac{3}{20} V_{\Phi}+\frac{M_1 V_1}{16}+\frac{M_2 V_1}{5 P e^{\Phi_{\infty}}} +\frac{M_1 V_2}{40 P e^{\Phi_{\infty}}} + O(M_i^4),\\[2mm]
\]
\[
f_{4,1}=-\frac{3}{10} M_1 V_1 + O(M_i^4), \qquad f_{4,0}=\frac{1}{10}V_{\Phi}-\frac{M_1 V_1}{2}+\frac{8 M_2 V_1}{15 P e^{\Phi_{\infty}}} -\frac{M_1 V_2}{10 P e^{\Phi_{\infty}}} + O(M_i^4),\\[2mm]
\]
\[
\Phi_{4,0}=-\frac{5}{4}V_{\Phi}-\frac{4 e^{\Phi_{\infty}}V_s}{3 P}+\frac{21}{2} M_1 V_1-\frac{16 M_2 V_1}{3 P e^{\Phi_{\infty}}} +\frac{8 M_2 V_2}{3 P^2 e^{2 \Phi_{\infty}}} -\frac{11 M_1 V_2}{2 P e^{\Phi_{\infty}}} + O(M_i^4),\\[2mm]
\]
\[
k_{4,2}=\frac{3}{2} P e^{\Phi_{\infty}} M_1 V_1, \qquad k_{4,1}=-\frac{3}{2}V_{\Phi} P e^{\Phi_{\infty}}+\frac{51}{8} M_1 V_1 P e^{\Phi_{\infty}}-2  M_2 V_1 -\frac{3}{2} M_1 V_2 + O(M_i^4),\\[2mm]
\]
\begin{equation}
g_{4,2}=k_{4,2}, \qquad g_{4,1}=k_{4,1},\qquad V_{\Phi}= k_{\Phi} (\frac{M_2 V_2}{P e^{\Phi_{\infty}}}+M_1 V_1).
\label{ksexpcoeff}
\end{equation}
where $k_{\Phi}$ scales in such a way that even when $M_{2,1} \to 0$, $V_{\Phi}$ remains finite.\\
All the other derived coefficients $\phi_{i,j}$ for $i \le 4 $ (and almost all for $i \ge 5$) are of order $(M_a)^i$ in the mass parameters. 

Once the IR and the UV expansions are found it is possible to relate the UV parameters to the IR ones, and find numerical solutions. Anyway plots are not particularly enlightening and we do not report them here.\
\subsection{The gluino condensate}
It was already pointed out in section \ref{glui} that the supergravity field dual to the gluino bilinear tr$\lambda \lambda$ is one of the polarization of $C_2=C_2^{RR}+i B_2^{NS}$.
Since
\begin{eqnarray}
H_3 = d B_2 &=& d u \wedge (\dot{g} g_1 \wedge g_2+\dot{k} g_3 \wedge g_4)+\frac{1}{2} (k-g) g_5 \wedge (g_1 \wedge g_3+g_2 \wedge g_4) \nn \\
F_3 = d C_2^{RR} &=& \dot{F} d u \wedge (g_1 \wedge g_3+ g_2 \wedge g_4)+ F g_5 \wedge g_1 \wedge g_2+ (2 P-F) g_5 \wedge g_3 \wedge g_4\nn
\end{eqnarray}
we have that $G_3 = d C_2$ behaves at $\t\to \infty$ as (we write only the polarizations along $T^{1,1}$):
\begin{displaymath}
\begin{array}{ccl}
G_3 &\simeq& \,\left[P+e^{-\frac{\t}{3}} (F_{1,1} \t +F_{1,0})+e^{-\t} (F_{3,5} \t^5+ \ldots+(P V_1 +O(M_i^3) )\t+\ldots)\right] g_5 \wedge g_1 \wedge g_2
\nn \\
&+& \, \left[P-e^{-\frac{\t}{3}}(F_{1,1} \t +F_{1,0})-e^{-\t} (F_{3,5} \t^5+ \ldots+(P V_1 +O(M_i^3) )\t+ \ldots)\right] g_5 \wedge g_3 \wedge g_4 \nn \\
&+& \,i \left[M_2 e^{-\frac{\t}{3}}+ e^{-\t}(k_{3,5} \t^5+ \ldots+ (P e^{\Phi_{\infty}} V_1+O(M_i^3)) \t+\ldots)\right]  g_5 \wedge (g_1 \wedge g_3+g_2 \wedge g_4)
\end{array}
\end{displaymath}
If we subtract the values of $G_3$ for two solutions with the same masses $M_i$ we get, at leading order (we relabel for simplicity $\Delta V_1$ as $V_1$):
\[
\Delta G_3 = (P V_1 \t e^{-\t}+\ldots) \ \omega_3, \quad \omega_3=-\left[g_5 \wedge(g_3 \wedge g_4 - g_1 \wedge g_2)+ i g_s g_5 \wedge (g_1 \wedge g_3+g_2 \wedge g_4) \right] 
\]
where $g_s= e^{\Phi_{\infty}}$.
Thus the polarization of $C_2$ we are interested in is
\[
C_2= - P V_1 \t e^{-\t} \omega_2, \qquad \omega_2=- \left[(g_1 \wedge g_3 + g_2 \wedge g_4)+ i g_s (g_1 \wedge g_2-g_3 \wedge g_4) \right]
\]
For large $\t$ we can perform the change of variables $\t \to \frac{1}{3}$ Log$(\epsilon^{-\frac{2}{3}} u)$.
Since the deformation parameter $\epsilon$ is related to the 4d mass scale as
$\epsilon \sim m^{-\frac{2}{3}}$, finally we get that for large $\t$
the operator we would like to associate with the gluino condensate scales as
\[
P V_1 \frac{m^3}{u^3} \texttt{Log}\frac{u^3}{m^3}
\]
This result can be expressed simply as $P V_1 m^3/u^3$, according to field theory prediction, with a redefinition of the scale as in \cite{dvlm,beme}; it would be interesting to evaluate the consequences of this issue.
\subsection{Vacuum energy}
Looking at the expression (\ref{potks}) for the effective action it is clear that, after integration by parts and the use of the equations of motion, the action becomes
\begin{equation}
I \sim \lim_{r \to \infty} \frac{3}{2} \dot{A}(r) e^{4 A(r)}
\end{equation}
In order to use the results of previous section we also have to perform the change of variables $d r \to d \t e^{4 p(\t)}$. The action is therefore given by:
\begin{equation}
I \sim \lim_{\t \to \infty} \frac{9}{2} \dot{A}(\t) e^{4 A(\t) -4 q(\t)+ 4 f(\t)}
\end{equation}
Even in the KS case, in order to cancel the divergences common to all vacuum energies, we have to subtract the actions of two solutions with different vevs. 
Again we take the supersymmetric solution, that is the KS one, as our  reference geometry.
It reads:
\begin{equation}
\begin{array}{cccccccc}
A_{KS}(\t)&=&\frac{1}{3} \t &+\frac{1}{6}\texttt{Log}(\t-1/4)&+&\frac{1}{6}\texttt{Log}(2^{-6} 3 P^2 e^{\Phi_{\infty}})
&+& O(e^{-2 \t}) \\
q_{KS}(\t)&=&  &+\frac{1}{6}\texttt{Log}(\t-1/4)&+&\frac{1}{6}\texttt{Log}(3 \sqrt{3} P^2 e^{\Phi_{\infty}})&+& O(e^{-2 \t})\\
f_{KS}(\t)&=& & & & & &O(e^{-2 \t})
\end{array}
\end{equation}
Accounting also for the freedom to shift $\t$ and $\Phi_{\infty}$ the action for this supersymmetric background is given, at leading order, by:
\begin{equation}
I_{KS} \propto \lim_{\t \to \infty} \;e^{\frac{4}{3} (\t+\t^*)}\;   \partial_{\t} \left(\frac{\t+\t^*}{3} +
\frac{1}{6} \log\left(\t-1/4+\t^*\right)+\frac{1}{6} \log (2^{-6} 3 P^2 e^{\Phi_{\infty}+\Phi^*}) \right)
\label{evks}
\end{equation}
The general solution have of course the form given in (\ref{UVKSpol}) and (\ref{UVKSexp}).\\
In KS solution $S_{\infty}$ is chosen in such a way that polynomial coefficients $a_{0,j}$, $q_{0,j}$ in (\ref{UVKSpol}) vanish. 
To account for the freedom to shift $s=g+k$  we take the value of $S_{\infty}$ for the general solution as the one that makes the above mentioned coefficients vanish, plus a correction $S^*$.
The polynomial series in $A(\t)$, $q(\t)$ thus becomes, at leading order in $S^*$:
\[
S^* \frac{1}{ 6 P \t e^{\Phi_{\infty}}} (1+\frac{1}{4 \t}+\frac{1}{(4 \t)^2}+\ldots) = S^* \frac{1}{6 P \t e^{\Phi_{\infty}}}\, \cdot \, \frac{1}{1-\frac{1}{4 \t}} = S^* \frac{1}{6 P e^{\Phi_{\infty}}(\t-\frac{1}{4})}
\]
So the action for the general solution is:
\begin{eqnarray}
I_{} \propto \lim_{\t \to \infty}  e^{4/3 \t+4 (A_{S}(\t)- q_{S}(\t)+f_{S}(\t))} \times
\partial_{\t}\left(\frac{\t}{3} +\frac{1}{6} \log (\t-\frac{1}{4})+ \frac{1}{6} \log (2^{-6} 3 P^2 e^{\Phi_{\infty}})
\right.\qquad \label{evgen}\\
\qquad \qquad \qquad \qquad \qquad \qquad \qquad \quad \left.
+\frac{S^*}{6 P e^{\Phi_{\infty}}(\t-\frac{1}{4})}+O\left((S^*)^2\right)+ A_{S}(\t) \right)\nn
\end{eqnarray}
Here $\phi_{S}(\t)$ indicates for each field its exponentially suppressed part.

To compare the energy of two solutions we have to match the corresponding metrics and scalar fields at the boundary.
Besides the metrics the two relevant fields are of course those related to the two gauge couplings, {\sl i.e.} $\Phi$ and $s=g+k$.\\
Using again the freedom to shift $\t$ and the constant values of $\Phi$ and $s$, the matching gives the following constraints:
\begin{displaymath}
\begin{array}{ccrcl}
\Phi(\t):&\qquad \quad &\Phi_{\infty}+\Phi^* &\quad \equiv \quad& \Phi_{\infty}+ \Phi_{S}(\t)\\
s(\t):&\qquad \quad & \; \; e^{\Phi_{\infty}+\Phi^*} P (\t+\t^*)- e^{\Phi_{\infty}+\Phi^*} P &\quad \equiv \quad& e^{\Phi_{\infty}} P \t- e^{\Phi_{\infty}} P+S^* + s_{S}(\t)\\
A(\t)-q(\t):& \qquad \quad &\frac{1}{3} (\t+\t^*) &\quad \equiv \quad& \frac{1}{3} \t+ (A_{S}(\t)- q_{S}(\t))
\end{array}
\end{displaymath}
Solving for infinitesimal $\t^*$, $S^*$ and $\Phi^*$ one gets:
\begin{eqnarray}
\Phi^* &\sim&  \Phi_{S}(\t) \nn \\
S^* &\sim& (-s_{S}(\t) + e^{\Phi_{\infty}} P \left((\t-1) \Phi_{s}(\t)+3 A_{S}(\t)-3 q_{S}(\t)\right)  \nn \\
\t^* &\sim&  3 (A_{S}(\t)- q_{S}(\t))
\label{star}
\end{eqnarray}
Substituting (\ref{star}) in (\ref{evks}) and subtracting (\ref{evgen}) we get, at leading order:
\begin{equation}
\Delta I \propto \lim_{\t \to \infty}  e^{4/3 \t} \left[\left(-8 f_S +
\frac{s_S}{ P e^{\Phi_{\infty}} \t^2} - \frac{3 p_s}{4 \t^2}-6 \dot{q}_S+\frac{\dot{s}_S}{P e^{\Phi_{\infty}} \t } +\frac{3 \dot{p}_S}{4 \t}\right)\left(1+O(\t^{-1})\right)\right]
\end{equation}
As in the MN case, if we use the full expression (\ref{UVKSexp}),(\ref{ksexpcoeff}) for the solutions, that is if we perform the calculation to all orders in the $M_a$, divergent terms of the form $\sim M^2 e^{2/3 \t}$ appear in the vacuum energy. They depend on the masses but not on the condensates, and reflect the fact that the vacuum energy in the softly broken theory is infinite.
All these terms however exactly cancel when computing difference
in the energy for different vacua.
The result is the same as if we do the calculation at first order in the mass parameters.
In both cases the leading surviving contributions come from $O(e^{-4/3 \t})$ and give a vacuum energy:
\[
\Delta I \propto Re( M_1 V_1 ) \t
\]
which, however, is still divergent.\\
Conversely, if we set $M_1$ to zero, contributions come from the next term and we have a finite result: 
\begin{equation}
\Delta I \propto e^{-\Phi_{\infty}} \left(Re ( M_2 V_2)+k'_{\Phi} Re(M_2 V_1 )\right) 
\end{equation}
It is not completely clear why giving a mass to both the fields associated with $y$ and $N_2$ leads to an infinite energy difference between two different vacua.
Nevertheless it is remarkable that at least in the particular case of only one mass parameter switched on we recover a finite vacuum energy.\newpage


\section{Conclusions}
We have studied softly broken $\mathcal{N}=1$ and $\mathcal{N}=2$ theories by deforming the Maldacena-Nu\~nez background and $\mathcal{N}=1$ theories by deforming the Klebanov-Strassler one.

We have found classes of solutions for both the systems that are globally regular. For these solutions we have computed the vacuum energy and verified that the degeneracy of supersymmetric vacua is lifted in the softly broken theory, according to expectation.
We note that even if SUSY is broken, the $U(1)_R$ symmetry is not, so many of the qualitative features of supersymmetric solutions persist. In particular we are still left with N vacua, which differ only in the phase of the gaugino condensate.
Moreover we remark that, at least for a subclass of backgrounds, energy differences between inequivalent vacua remain finite.

The key ingredient involved either in the singularity resolution, either in vacuum computations is for both systems the gaugino condensate.
An intriguing aspect is that information about condensates and vacuum energies are encoded in the subleading UV behavior of the supergravity solutions, but we have seen that UV parameters are actually fixed by regularity and boundary conditions in the IR.
In other words, supergravity solutions show manifestly the deep interplay between UV and IR phenomena. Further investigations in that direction would be very interesting.

A correlated problem concern the analysis of other features of the softly broken theory encoded in the full solutions or in their IR behavior, such as for example the glueball spectrum. 

Different kind of deformations of the 5D part of backgrounds, for example with suitable operators, analogue to the ones of \cite{eva} for AdS, would be another fruitful direction of analysis.

We should also discuss the issue of stability of general solutions, since supersymmetry is not protecting them any more. It is generally assumed that, since $\mathcal{N}=1$ theory has a mass gap, at least for small deformations stability is preserved. A more detailed analysis is nevertheless necessary to exclude the possibility of tachionic instabilities.

Finally, as pointed out in \cite{gubser1}, globally regular solutions with finite vacuum energy, such as the ones of MN setup, have four dimensional Poincare invariance and positive cosmological constant and could be useful in string cosmology. For example, in \cite{gubser1} was outlined a possible model for Universe evolution that could account for cosmological supersymmetry breaking.

\section*{Acknowledgments}
The author would like to thank first of all Prof. A . Zaffaroni, who suggested this investigation, for his continuous help.
He would also like to thank Dott. A. Cotrone and Dott. F. Bigazzi for useful discussions and the Physics Department of the Milano-Bicocca University for kind hospitality.
The author is partially supported by INFN and MURST.

\end{document}